\documentclass[twocolumn]{aastex62}
\usepackage{amssymb}
\usepackage{color}
\usepackage{amsmath}
\usepackage{tablefootnote}
\usepackage{comment}
\usepackage{lipsum}
\usepackage{hyperref}
\usepackage{soul}
\usepackage{graphicx}
\usepackage{natbib}
\usepackage{comment}
\shortauthors{Broussard et al.}
\shorttitle{Galaxy Star Formation Stochasticity}

\revised{January 4, 2019}
\submitjournal{Astrophysical Journal}

\begin{document}

\title{Star Formation Stochasticity Measured from the Distribution of Burst Indicators}

\author{Adam Broussard}
\affiliation{Department of Physics and Astronomy, Rutgers, The State University of New Jersey, 136 Frelinghuysen Rd, Piscataway, NJ 08854, USA}
\email{adamcbroussard@physics.rutgers.edu}

\author{Eric Gawiser}
\affiliation{Department of Physics and Astronomy, Rutgers, The State University of New Jersey, 136 Frelinghuysen Rd, Piscataway, NJ 08854, USA}
\affiliation{Center for Computational Astrophysics, Flatiron Institute, 162 5th Ave, New York, NY 10010, USA}
\email{gawiser@physics.rutgers.edu}

\author{Karthiek Iyer}
\affiliation{Department of Physics and Astronomy, Rutgers, The State University of New Jersey, 136 Frelinghuysen Rd, Piscataway, NJ 08854, USA}
\email{iyer@physics.rutgers.edu}

\author{Peter Kurczynski}
\affiliation{Department of Physics and Astronomy, Rutgers, The State University of New Jersey, 136 Frelinghuysen Rd, Piscataway, NJ 08854, USA}
\email{pkurczynski@mac.com}

\author{Rachel S. Somerville}
\affiliation{Center for Computational Astrophysics, Flatiron Institute, 162 5th Ave, New York, NY 10010, USA}
\affiliation{Department of Physics and Astronomy, Rutgers, The State University of New Jersey, 136 Frelinghuysen Rd, Piscataway, NJ 08854, USA}
\email{rsomerville@flatironinstitute.org}

\author{Romeel Dav\'e}
\affiliation{Department of Physics and Astronomy, University of the Western Cape, Bellville, Cape Town 7535, South Africa}
\affiliation{South African Astronomical Observatories, Observatory, Cape Town 7925, South Africa}
\affiliation{African Institute for Mathematical Sciences, Muizenberg, Cape Town 7945, South Africa}
\affiliation{Institute for Astronomy, Royal Observatory, Edinburgh EH9 3HJ, UK}
\email{romeeld@gmail.com}

\author{Steve Finkelstein}
\affiliation{Department of Astronomy, The University of Texas at Austin, Austin, TX 78712, USA}
\email{stevenf@astro.as.utexas.edu}

\author{Intae Jung}
\affiliation{Department of Astronomy, The University of Texas at Austin, Austin, TX 78712, USA}
\email{itjung@astro.as.utexas.edu}

\author{Camilla Pacifici}
\affiliation{Space Telescope Science Institute, Baltimore, MD 21218, USA}
\affiliation{Goddard Space Flight Center, Code 665, Greenbelt, MD 20771, USA}
\email{cpacifici@stsci.edu}

\begin{abstract}
One of the key questions in understanding \deleted{the}\added{galaxy} formation and evolution \deleted{of galaxies }is how starbursts affect the assembly of stellar populations in galaxies over time.  We define a burst indicator ($\eta$), which compares a galaxy's star formation rates on short ($\sim10$\,Myr) and long ($\sim100$\,Myr) timescales.  To estimate $\eta$, we apply the detailed \deleted{temporal luminosity response}\added{time-luminosity relationship} \deleted{of}\added{for} H$\alpha$ and near-ultraviolet emission to simulated star formation histories (SFHs) from Semi-Analytic Models and the \textsc{Mufasa} hydrodynamical cosmological simulations.  \deleted{We find that t}\added{T}he average \deleted{value }of $\eta$ is not\deleted{ itself} a good indicator of star formation stochasticity (burstiness); indeed\added{,} we show that this average should be close to zero unless the\added{ galaxy} population \deleted{being studied }has an average SFH \deleted{which}\added{that} is rising or falling rapidly.  Instead, \deleted{analyzing }the width of the \added{$\eta$ }distribution\deleted{ of $\eta$ for ensembles of galaxies allows us to} characterize\added{s} the \deleted{level of }burstiness \deleted{in}\added{of} a galaxy population's recent star formation\deleted{ history}.  We find this width to be robust to variations in stellar initial mass function (IMF) and metallicity.  We apply realistic noise and selection effects to the models to generate mock \textit{HST} and \textit{JWST} galaxy catalogs and compare these catalogs with 3D-HST observations of 956 galaxies at $0.65<z<1.5$ detected in H$\alpha$.  Measurements of $\eta$ are unaffected by dust measurement errors under the assumption that $E(B-V)_\mathrm{stars}=0.44\,E(B-V)_\mathrm{gas}$ (i.e. $Q_\mathrm{sg}=0.44$).  However, setting $Q_\mathrm{sg}=0.8^{+0.1}_{-0.2}$ removes an unexpected dependence of the average value of $\eta$ upon dust attenuation and stellar mass in the 3D-HST sample while also resolving disagreements in the distribution of star formation rates.  However, even varying the dust law cannot resolve all discrepancies between the simulated and the observed galaxies.
\end{abstract}


\section{Introduction}

The history of estimating SFRs from galaxy observables dates back at least to the identification of the Hubble classification of galaxies; for a review of this history, the reader is referred to \cite{Kennicutt1998}.  In particular, starbursts in galaxies have been extensively studied over the past several decades \citep{McKee2007, Dale2005, Baldry2004, Brinchmann2004, Baldry2006, Kauffmann2003, Tan2001, Gavazzi1996, Tomita1996, Kennicutt1994, Bothun1990}.  One aspect of this subject which has gained increased attention in recent years is the difference in timescales of star formation to which various SFR estimators may be sensitive.  Such differences become crucial in assessing whether star formation has occurred as a gradual process over time, or whether it has occurred suddenly in burst(s).  Contemporary stellar population synthesis methods enable a more complete assessment of these differences than was possible in the past.

Several papers have been published in recent years investigating whether galaxies tend toward short starburst and quenching periods, or relatively continuous star formation (e.g. \citealt{Hopkins2014, Shen2014, Dominguez2015, Sparre2015, Guo2016, Sparre2017}).  Efforts have been made both to understand the timescales over which a galaxy's star formation rate (SFR) undergoes periods of starburst, and to quantify the magnitude of variations in the SFR consistently.  Accurate measurements of these quantities would give valuable insight into the physical processes governing short-timescale changes in SFR such as feedback from supernovae or active galactic nuclei (AGN).  For example, in the FIRE simulations, which resolve the GMC-scale structure of the ISM and explicitly treat stellar feedback (from supernovae, radiation pressure on dust grains, wind and photoheating), the clustered nature of star formation (both in time and space) and resulting impulsive nature of stellar feedback result in bursty star formation histories, especially in lower-mass galaxies (Sparre at al. 2017). However, these simulations do not include various other potentially relevant feedback processes, such as cosmic rays and AGN, which may alter the time variability of the SFR. Variations in gas inflow rates also drive variations in the SFR, but these tend to occur on longer timescales than feedback-induced burstiness (Sparre et al. 2015).  Accurate measurements of galaxy SFR \deleted{stochasticity}\added{variability} could reveal the degree to which intrinsic scatter in the SFR-M$_*$ relationship (e.g., \citealt{Shivaei2015, Kurczynski2016, Matthee2018}) is due to variations in SFR, as opposed to galaxies which systematically form stars more or less efficiently than their peers of equal mass over long ($> 100$\,Myr) timescales.

Two commonly used tracers of a galaxy's SFR are H$\alpha$ and the ultraviolet (UV) continuum.  H$\alpha$ emission arises from recombination of gas ionized primarily by O-type stars, and therefore traces timescales on the order of the lifetime of such stars ($\lesssim 5$ Myr).  The UV continuum primarily is emitted by O- and B-type stars, which have lifetimes of $\lesssim 100$ Myr.  These two measurements allow us to probe the immediate and recent SFR for a galaxy, which can help determine if it is actively bursting or quenching.

A common method of quantifying burstiness is to use the average UV to H$\alpha$ flux ratio.  An analysis of galaxy star formation burstiness by \cite{Weisz2012} makes use of a toy model with periodically repeating star formation bursts, characterized by a duration, a period, and an amplitude.  They find that observed flux ratio distributions of 185 \textit{Spitzer}-observed galaxies can be fit with these toy models without invoking other physical processes.  They note a decrease in H$\alpha$ to UV flux ratio and increase in scatter for lower mass galaxies in their sample.  \cite{Iglesias2004} find that the standard deviation of the ratio of H$\alpha$ to UV flux could be indicative of variable SFR, but also notes that variable amounts of metals in the interstellar medium (ISM), variable initial mass function (IMF), or absorption of Lyman continuum photons could have similar effects.  Additionally, \cite{Guo2016} finds a decrease in the H$\beta$-to-FUV ratio with decreasing galaxy mass and notes that bursty star formation on timescales of a few tens of Myr is a plausible explanation for this variation, and that with decreasing $\mathrm{M_*}$, burstiness is increasingly necessary to reproduce the expected level of variation.

A ratio of long-timescale SFR to short-timescale SFR (i.e., of continuum flux density to emission line flux) is often used to probe burstiness \citep{Sullivan2000, Iglesias2004, Boselli2009, Wuyts2011, Dominguez2015, Guo2016, Sparre2017}.  Some authors \cite{Dominguez2015} (see their Figure 6) noted a significant benefit to using this ratio because in highly variable SFHs, the 10\,Myr star formation rate (SFR$_{10}$) can effectively become zero, generating a population of high outliers.  It is unclear, however whether typical observational detection limits allow this theoretical advantage to be realized.  This motivates a thorough study of the performance of potential burstiness estimators in the context of existing and anticipated observational capabilities.  We seek a definition for burstiness which is formulated from intrinsic galaxy SFRs and can be measured from commonly available observables.  We start with a mathematical framework for quantifying rapid changes in a galaxy's SFR, which we can then apply to ensembles of galaxies in order to determine a population burstiness value.  In the end, this methodology results in an intuitively understandable burstiness estimator with a firm grounding in theory that remains practically applicable.  We perform this study in the redshift range $0.8<z<1.2$ for ease of comparison with data from 3D-HST.

In Section \ref{Data}, we present the data we use for our analysis, including both simulated and observed data. Section \ref{Sec: Methodology} contains a discussion of the process of developing the burst indicator from theory, as well as transitioning it into an \deleted{observed}\added{observational} estimator.  Additionally, we discuss methods of analysis which can be used to determine the burstiness of a galaxy population at a particular epoch.  Section \ref{Sec: IMF, metal, SFH} shows the effects of variations in the high-mass IMF slope, galaxy metallicity, and bulk trends in star formation history (SFH) on measurements of population burstiness.  In Section \ref{Sec: Mocks}, we generate a mock catalog of galaxies from the simulations, and we apply realistic Hubble Space Telescope (HST) and James Webb Space Telescope (JWST) selection and noise effects to examine their effects on burstiness measurements.  Section \ref{Sec: Gas, Star Dust} shows the relationship between $A_{V,\mathrm{stars}}$ and $A_{V,\mathrm{gas}}$ in the mock catalogs and in the 3D-HST data, and explains the potential and limitations of assuming that galaxy burstiness is independent of dust prescription.  Section \ref{Sec: Results} summarizes the results of the mock catalogs by comparing them to the 3D-HST data and demonstrating improvements which will be made when this analysis can be performed on observed spectra from JWST.  Finally, Section \ref{Sec: Conclusions} summarizes our methods, analysis, and results.

\section{Data}\label{Data}

\subsection{Simulations}\label{Sec: Sims}

Throughout this work, we make use of a large-volume hydrodynamical simulation ({\sc Mufasa}, \citealt{Dave2017}) as well a as semi-analytic model (SAM, \citealt{Somerville2008}).  The {\sc Mufasa} hydrodynamical simulations attempt to model the fluid dynamics of the ISM, intergalactic medium (IGM), and intercluster medium (ICM) in and around galaxies.  This type of simulation has the benefit of attempting to directly simulate many of the physical processes involved with gas flows in galaxies, with the drawback of being computationally expensive.  {\sc Mufasa} calculates star formation rates based on the molecular gas content of individual gas fluid elements using a \cite{Schmidt1959} law, where the molecular gas fraction for each fluid element is computed using a sub-resolution prescription based on \cite{Krumholz2011}.  An artificial pressure is added to star-forming gas in order to ensure that the Jeans mass is always resolved, which for this work is relevant because it smooths the ISM and thus is expected to damp short-term fluctuations in the star formation rate.  Furthermore, owing to its nominal spatial resolution of 0.5 kpc/h, {\sc Mufasa} (and similar cosmological simulations) cannot accurately model the strong compactification associated with merger-driven starbursts.  These resolution limitations suggest that relatively low resolution cosmological simulations such as {\sc Mufasa} may underestimate the burstiness in smaller and more extreme starbursting systems \cite{Sparre2015}.  Overall, however, extreme starbursts are rare and thus expected (and observed) to contribute only a small fraction to global galaxy growth.

Star formation histories in the {\sc Mufasa} simulations are represented by instantaneous bursts of star formation in which a given amount of stellar mass is formed.  In order to convert these delta functions into a continuous star formation history, we convolve them with an Epanechnikov kernel \cite[pg. 255]{Ivezic2014} with a width of 50\,Myr.  This gives an approximate half-width-half-maximum of $\sim10\,\mathrm{Myr}$ such that the time resolution of the {\sc Mufasa} SFHs is roughly similar to that of the SAM SFHs.  In addition, we require that a given galaxy has undergone at least 128 star formation events (with a minimum star particle mass of $\sim10^{6.5}M_\odot$) in order for the star formation history to be considered reliable, so this is the only cut applied to the simulation for our sample.  This leaves us with star formation histories for $1461$ {\sc Mufasa} galaxies.  

Instead of attempting to directly solve the physical equations, the semi-analytic models achieve greater computational efficiency by applying simplified but physically motivated recipes to underlying dark matter structure.  Dark matter halos are traced through N-body simulations, culminating in a dark halo merger history.  Because galaxies form within these dark matter halos, empirically motivated recipes pertaining to their gas content, star formation, and other properties can be applied to determine their physical properties \citep{Somerville2015}.  We apply a mass cut of $\mathrm{M_*} > 10^8 \mathrm{M}_\odot$ and randomly sample $10^4$ galaxy SFHs from the SAM catalog provided due to computational constraints.  Star formation histories for the SAMs are provided as measurements of star formation in 10\,Myr time intervals although the underlying time step can be shorter than this.

The SAM star formation histories are calculated using two star forming modes: a normal quiescent mode for isolated discs and a merger-driven starburst mode \citep{Somerville2008}.  Quiescent star formation utilizes a recipe based on the empirical Schmidt-Kennicutt law \citep{Kennicutt1998, Kennicutt1989} in which the star formation rate surface density in the galactic disk has a power law dependence on the gas surface density when the gas surface density is above a threshold:

\begin{align*}
\Sigma_\mathrm{SFR}=A_\mathrm{Kenn}~{\Sigma_\mathrm{gas}}^\mathrm{N_K}
\end{align*}
where $A_\mathrm{Kenn}=1.67\times 10^{-4}$ and $\mathrm{N_K} = 1.4$.  $\Sigma_\mathrm{gas}$ is in units of $\mathrm{M}_\odot~\mathrm{pc^{-2}}$ and $\Sigma_\mathrm{SFR}$ is in units of $\mathrm{M}_\odot~\mathrm{yr^{-1}~pc^{-2}}$.  The merger-driven mode is based on the results of hydrodynamic simulations of galaxy mergers in \cite{Robertson2006} and \cite{Hopkins2009}.  The SAM accounts for the feedback of supernovae, massive stars, and AGN, as well as effects of gas recycling, metallicity, and loss of angular momentum in mergers.  For more details on the model, see \cite{Somerville2008}, \cite{Somerville2012}, and \cite{Porter2014}.

Global trends such as the buildup of stellar mass over cosmic time, SFR-M$_*$ relationship, and galaxy number densities are all well predicted by these simulations up to $z\sim4$.  Galaxy morphologies are also reproducible using these simulations although it is not yet clear how well these reproduce observed morphology distributions or the details of their evolution.  In addition, these models correctly predict qualitative trends in the fraction of star forming and quiescent galaxies as well as relationships between star formation and morphology.  One relevant limitation of these simulations is the prediction of self-similar SFHs while observations indicate a stronger mass dependence.  An detailed overview of the tools involved with modeling galaxy evolution along with a comparison of these another models to observations can be found in \cite{SomervilleDave2015}.

{\sc Mufasa} provides us with the full SFH of each simulated galaxy, but due to computational limitations, the SAM set of SFHs are limited to $z>0.8$.  This limitation has no effect on our analysis in this work that is applied to $0.8<z<1.2$.  Both \textsc{Mufasa} and the SAMs have a time resolution close to 10\,Myr, and our analysis is unchanged by slight changes in time sampling.  We sample the time axis of each simulation at 1\,Myr intervals for consistency across the sample, but as $\mathrm{SFR_{H\alpha}}$ samples timescales somewhat shorter than 10\,Myr, it would be advantageous to repeat this analysis in the future using models with even higher time resolution.  We exclude the first 100\,Myr of each galaxy's star formation history, when the 100\,Myr timescale SFR is not well defined.  These SFHs allow us to form a theoretical indicator of star formation burstiness directly from physical properties of each galaxy, without having to work backwards from observable quantities.

Flexible Stellar Population Synthesis (FSPS, \citealt{Conroy2009, Conroy2010}) is a software package designed to simulate the spectra of stellar populations.  In particular, it combines calculations of stellar evolution with stellar spectral libraries to produce spectra for mono-metallic, coeval simple stellar populations.  Throughout this work, we implement parameters in FSPS to generate stellar spectra that: utilize a Chabrier IMF (except in Section \ref{Fig: var metal IMF} where indicated), include nebular line and continuum emission, implement a \cite{Calzetti2000} dust law, and have $Z = 0.2\,Z_\odot$, matching the SpeedyMC fits to observed 3D-HST data and in agreement with \cite{Gao2018}.  Nebular emission is calculated using the FSPS implementation of the photoionization code, CLOUDY \citep{Byler2017}, which assumes a constant density spherical shell of gas surrounding the stellar population.  Spectra are calculated at 1\,Myr time intervals over a period of 3\,Gyr.

\subsection{Observations}

We utilize 3D-HST data \citep{Skelton2014} as a comparison against our simulated data.  3D-HST was an HST Treasury program (Programs 12177 and 12328; PI: P. van Dokkum) that covered five extragalactic fields with grism spectroscopy.  The survey is described in \cite{Brammer2012a}.  

A subset of galaxies in the 3D-HST catalog were chosen for estimating H$\alpha $ fluxes from grism observations and photometric Spectral Energy Distribution (SED) fitting to determine other physical parameters.  The sample was selected to have SNR$_\mathrm{H\alpha} > 10$ to ensure acceptable data quality as determined by visual inspection of grism data for representative galaxies in the catalog.  Additional 3D-HST grism catalog quality criteria included \texttt{use\_zgrism = 0}, \texttt{use\_phot = 1} and restricting the redshift range to $0.65 < z < 1.50$ for physical consistency with G141 grism detection of H$\alpha$.  Further SED-fit quality criteria were imposed as discussed below.

Photometry from the catalog of \cite{Skelton2014} were used for SED fits.  Fitting was performed with the SpeedyMC software  \citep{Acquaviva2011, Acquaviva2012}.  In brief, SEDs were fit to \cite{Bruzual2003} population synthesis models, assuming a \cite{Salpeter1959} IMF, a linear-exponential star formation history, the dust prescription from \cite{Calzetti2000}, and $\log_{10}\left(Z/Z_\odot\right) = 0.2$; details of the SED fitting procedure are presented in \cite{Kurczynski2016}.  Results of the SED fitting included estimates of stellar masses, SFRs, $E(B-V)_\mathrm{stars}$ and star formation history timescale ($\tau$) values for each galaxy in the sample.  SED fit quality criteria restricted the sample to sources with \cite{Gelman1992} GR convergence parameter $\mathrm{GR} < 0.2$ and reduced $\chi^2 < 3.0$.

To investigate AGN contamination in the sample, galaxies in the GOODS-S field were position-matched against x-ray catalog of \cite{Xue2011}, (see also \citealt{Kocevski2017}).  7/264 sources in GOODS-S were found to be consistent with an AGN.  Their estimated masses and SFRs tended to be greater than the sample averages, but they were not exceptional outliers.  Since X-ray data of this depth are not available in the other fields, we estimate an AGN contamination fraction of 3\% in our analyses, and we do not expect that AGN significantly bias our results.

For each galaxy, H$\alpha$ flux was estimated from the cataloged grism flux values assuming a constant [NII] fraction of 0.1.  An intrinsic (i.e. dust corrected) H$\alpha$ flux estimate was obtained from the observed values using SED-based $E(B-V)_\mathrm{stars}$ estimates as follows:  the estimated $E(B-V)_\mathrm{stars}$ values and their uncertainties were used in conjunction with a Calzetti attenuation law to estimate the dust attenuation at H$\alpha$ wavelength for each galaxy. The \cite{Calzetti2000} attenuation prescription used $R_v = 4.05$ and $\mathrm{E(B-V)_\mathrm{stars}} = 0.44 \mathrm{E(B-V)_\mathrm{gas}}$ throughout.  Although a subset of galaxies had H$\beta$ detections reported in the 3D-HST catalog, Balmer decrement dust correction was found to be unreliable for these galaxies.  

3D-HST grism data were obtained by using a detection in the $F140W$ as a direct image, which serves to establish the calibration wavelength of the spectra.  Detections in the G141 grism have an average 5$\sigma$ sensitivity of $f = 5.5\times10^{-17}~\mathrm{erg~s^{-1}~cm^{-2}}$ \citep{Brammer2012a}.  After applying these cuts, we are left with 956 galaxies in our catalog, which we utilize in this work.

\section{Burstiness Methodology}\label{Sec: Methodology}

It is important to note that it is observationally difficult to determine the burstiness of an individual galaxy.  We may be able to determine if a galaxy is currently undergoing a starburst or quenching episode, but the detailed rises and falls in its recent SFH are not typically measurable.  As a result, it is necessary instead to measure the burstiness of an ensemble of galaxies.  For a given population of galaxies, we can analyze the distribution of their current states - bursting, quenching, or continuously star forming - to gain insight into the degree of star formation burstiness for the ensemble.

\begin{figure*}
\centering
\includegraphics[width = \textwidth]{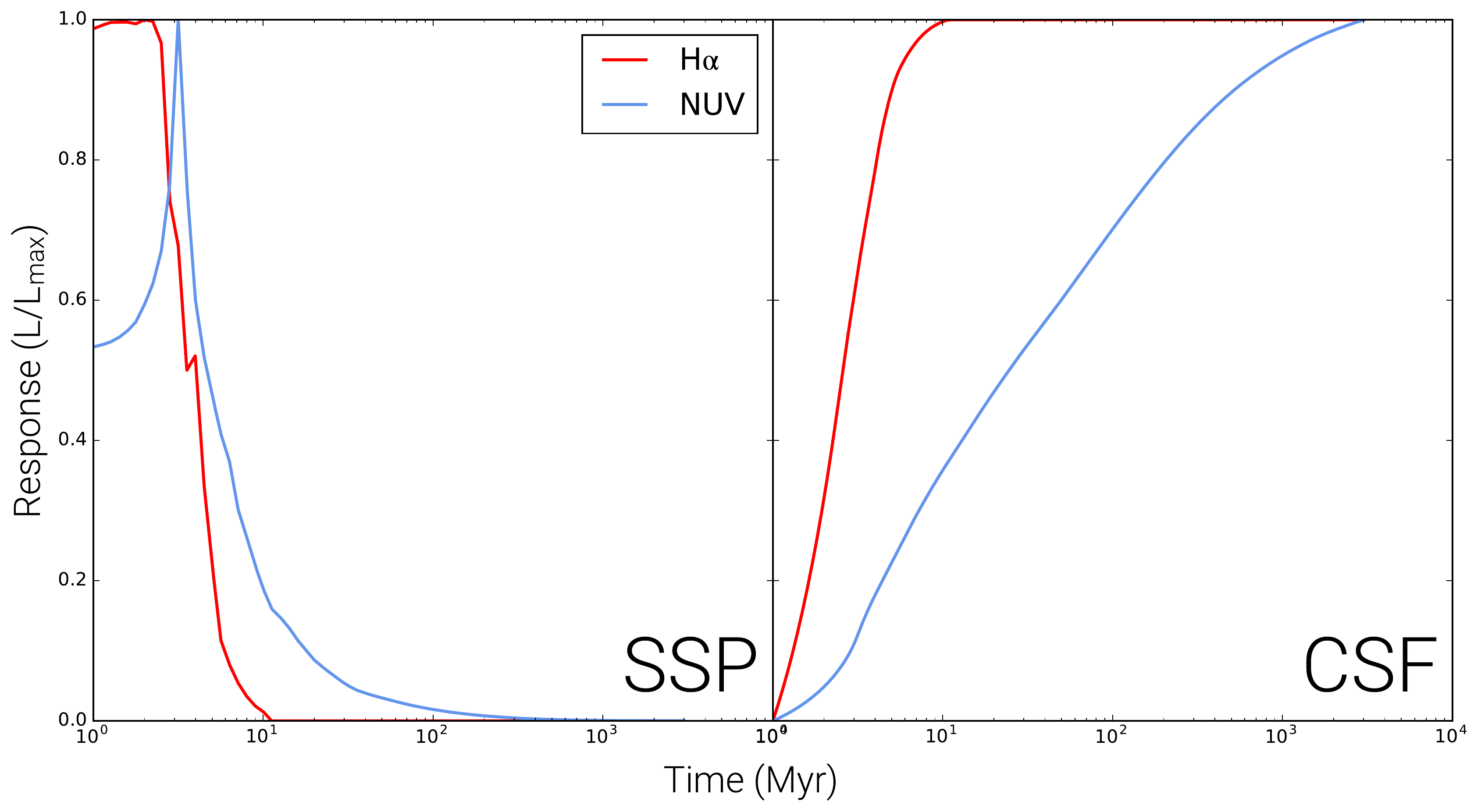}
\caption{\label{Fig: Response}Normalized luminosity \deleted{responses}\added{as a function of time} for relevant star formation tracers shown as a function of time, as generated by FSPS.  The left panel shows the response of each tracer to a simple stellar population (SSP) while the right panel shows the response of each tracer in a continuous star formation (CSF) case.  The luminosity of each tracer has been normalized so that the maximum is 1.  As is shown in the right panel, the time at which H-$\alpha$ emission reaches [10\%, 50\%, 90\%] of its maximum value is [1.32, 2.58, 5.03] Myr.  The time at which NUV emission reaches [10\%, 50\%, 90\%] of its maximum value is [2.87, 24.56, 520.64] Myr.}
\end{figure*}

\subsection{Defining a Burst Indicator ($\eta$)}

The first task in defining galaxy star formation burstiness is to determine the state of an individual galaxy: is it currently in a period of starburst, is it forming stars at a relatively constant rate, or is it quenching?  We will call our quantitative measurement of this the burst indicator (represented by $\eta$) because it describes the current state of the galaxy as opposed to its longer term SFH.

In order to determine the burst state of a galaxy, we first must be able to separate the continuous star formation of a given SFH from the bursting and quenching periods.  Several previous works have indicated that periods of starburst are likely shorter than $\sim$100\,Myr \citep{Dominguez2015,Guo2016}.  As a result, we use the median SFR over a  period of 100\,Myr to represent the continuous SFR, while using the SFR averaged over 10\,Myr to represent the short timescale star formation.  Utilizing a median for the longer timescale SFR gives better sensitivity to short-timescale bursts or dips in star formation for our burstiness estimator.  We will call the 100\,Myr median star formation rate $\mathrm{\widetilde{SFR}}_{100}$ and the 10\,Myr star formation rate SFR$_{10}$.

We can now use these star formation rate measurements to form a theoretical burst indicator.  Here, we mean theoretical in the sense that the incorporated quantities are intrinsic star formation rates (i.e. they are not easily observable using star formation tracers).  This theoretical burst indicator represents the degree to which a galaxy is in a period of rising, declining, or constant star formation.  We define the theoretical burst indicator, $\eta_\mathrm{th}$, as follows\footnote{Other definitions are clearly possible; we arrived at the choice of $\log_{10}(\mathrm{SFR_{10}/SFR_{100}})$ in order to enable later comparison with \deleted{observed}\added{observational} \deleted{SFR}\added{burstiness} estimators and because it performs well for both types of model SFHs considered here.  We found no significant differences in results between this definition and a simple ratio, $\mathrm{SFR_{10}/\widetilde{SFR}_{100}}$, or its inverse.}

\begin{align}\label{Eqn: eta_th}
\eta_\mathrm{th}=\log_{10}(\text{SFR}_{10}/\mathrm{\widetilde{SFR}}_{100})
\end{align}

This ratio is comparable to ones used to investigate burstiness in other works (see \citealt{Guo2016}, \citealt{Weisz2012}, and references therein).  Because $\eta_\mathrm{th}$ is a logarithm of a ratio of SFRs, the numerator and denominator are equally weighted.  When the $\eta_\mathrm{th}$ as defined here is $>0$, the galaxy will be in a period of increasing SFR (e.g., starburst) because the shorter-timescale SFR is greater than the longer-timescale SFR.  Conversely, $\eta_\mathrm{th}<0$ corresponds to a period of decreasing SFR (e.g., quenching).  Finally, $\eta_\mathrm{th} \approx 0$ corresponds with approximately constant star formation over 100\,Myr timescales.

\subsection{Theoretical and Observational Burst Indicators}

The theoretical burst indicator $\eta_\mathrm{th}$ uses quantities, SFR$_{10}$ and $\mathrm{\widetilde{SFR}}_{100}$ that are not available to us when observing galaxies, so we now define an \deleted{observed}\added{observational} burst indicator ($\eta$) based on measurable values and determine the relationship between $\eta_\mathrm{th}$ and $\eta$.  This process of starting with a theoretical burst indicator and transitioning to \deleted{observed}\added{observational} values is similar to that of \cite{Sparre2017}.  This \deleted{observed}\added{observational} burst indicator has some similarities to the birthrate parameter $b$ from \cite{Scalo1986}, with the primary difference being that the denominator of $b$ is a disk-averaged SFR while ours probes the recent star formation history.

Here, we will use the $\mathrm{H\alpha}$ star formation rate (SFR$_\mathrm{H\alpha}$) and the near-ultraviolet star formation rate (SFR$_\mathrm{NUV}$; $\mathrm{ 2700\,\AA < \lambda < 2800\,\AA}$) as our proxies for the short and long timescale star formation rates we defined for $\eta_\mathrm{th}$.  We can then define the \deleted{observed}\added{observational}\deleted{ equivalent of the} burst indicator: 

\begin{align}\label{Eqn: eta}
\eta & = \log_{10}(\mathrm{SFR_{H\alpha} / SFR_{NUV}}).
\end{align}

While it is true that the dust-corrected H$\alpha$ flux and NUV flux density could be used in the definition of $\eta$ rather than their corresponding SFRs, we find that it is more intuitive to use the definition in Equation \ref{Eqn: eta} because $\eta=0$ for constant SFR, and because it is more consistent with our definition of $\eta_\mathrm{th}$ in Equation \ref{Eqn: eta_th}.  We note that SFR$_{10}$ is more similar to SFR$_\mathrm{H\alpha}$ than $\mathrm{\widetilde{SFR}}_{100}$ is to SFR$_\mathrm{NUV}$.  This is because SFRR$_{10}$, SFR$_\mathrm{H\alpha}$, and SFR$_\mathrm{NUV}$ are effectively weighted arithmetic means while $\mathrm{\widetilde{SFR}}_{100}$ is a 100\,Myr median.  It is sometimes practical to make use of other tracers such as SFR$_\mathrm{H\beta}$ for the short timescale and SFR$_\mathrm{FUV}$ or SFR$_\mathrm{IR}$ for the long timescale star formation rate, but this process is easily repeated for those quantities.  We generate simple stellar population (SSP) and continuous star formation (CSF) histories and use FSPS \citep{Conroy2009} to compute the luminosity \deleted{responses} as a function of time in the H$\alpha$ line \deleted{luminosity}\added{emission} and broadband photometry.  This allows us to directly estimate the timescales probed by various tracers.  This usage is important because the NUV tracer, for example, takes more than 10 times as long to reach 90\% of its maximum luminosity as it takes to reach 50\% of the maximum, making any single timescale a poor description of the\added{ time-dependent} luminosity \deleted{response}.  We also consider the potential for binary evolution to affect the timescales to which star formation tracers are sensitive.  FSPS includes a functional implementation of the Binary Population and Spectral Synthesis code (BPASS; \citealt{Eldridge2017,Stanway2018}), which models the effects of the evolution of close stellar binaries on observed spectra.  For $Z = 0.2\,Z_\odot$ used in this analysis, we found the \added{time-dependent }luminosity \deleted{response} of both H$\alpha$ and the NUV to be effectively identical to those calculated for single stellar systems, indicating that stellar binaries do not have a significant effect on tracer timescales in our analysis.

\begin{figure*}[t]
\centering
\includegraphics[width = \textwidth]{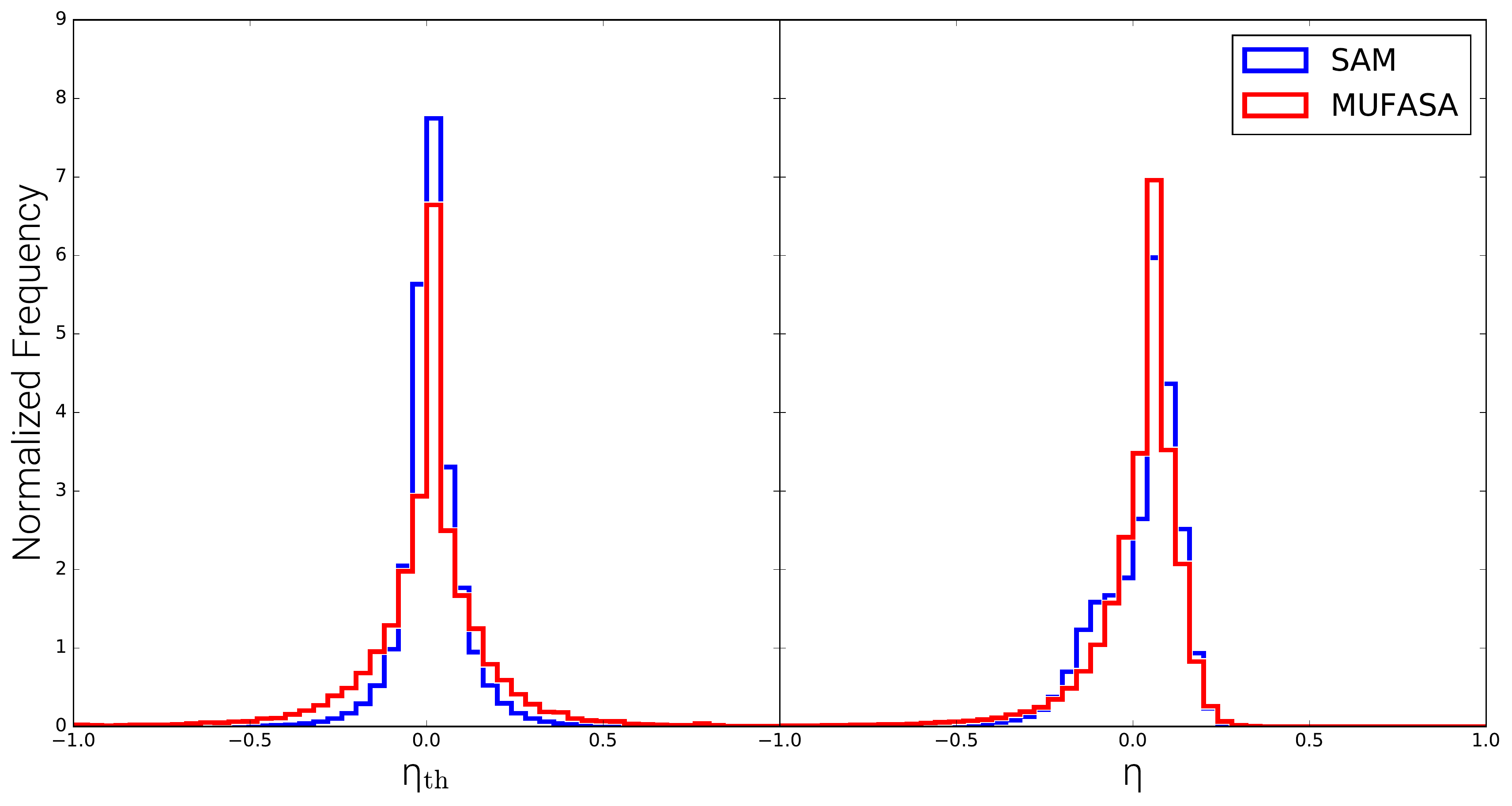}
\caption{\label{Fig: BSI}Histograms showing the distribution of both the theoretical burst indicator ($\eta_\mathrm{th}$; left) and the \deleted{observed}\added{observational} burst indicator ($\eta$; right).  The distribution for the {\sc Mufasa} sample is shown in each panel in red while the distribution for the SAM sample is shown in blue.  Each sample is limited to $z>0.8$ for consistency.  Integrated frequency has been normalized to 1 for ease of comparison between the differently-sized samples.  The distribution of $\eta$ is obtained by applying the normalized \added{relationship between }luminosity\added{ and time} \deleted{responses} shown in Figure \ref{Fig: Response} to the star formation histories to obtain $\mathrm{SFR_{H\alpha}}$ and $\mathrm{SFR_{NUV}}$ (see Equation \ref{Eqn: eta}).  The statistics for this figure are listed in Table \ref{Tbl: simstats}.}
\end{figure*}

\begin{deluxetable}{crlll}
\tablecaption{Burstiness Statistics}
\vspace{-0.2cm}
\tablehead{  &  & \colhead{$\mu$} & \colhead{$\sigma$} & \colhead{$\gamma$}}
\startdata
$\eta_\mathrm{th}$ & {\sc Mufasa} & \phantom{-}0.00 & 0.21 & -1.56 \\
 & SAM & \phantom{-}0.01 & 0.10 & 0.36\\ \hline
$\eta$ & {\sc Mufasa} & -0.02 & 0.24 & -10.40\\
 & SAM & \phantom{-}0.01 & 0.15 & -8.07\\
\enddata
\tablecomments{\label{Tbl: simstats}The average ($\mu$), standard deviation ($\sigma$), and skewness ($\gamma$) for $\eta$ and $\eta_\mathrm{th}$ measured in the full sample of galaxies from the SAM and {\sc Mufasa} simulations.  These statistics correspond with the histograms shown in Figure \ref{Fig: BSI}.}
\end{deluxetable}

Figure \ref{Fig: Response} shows the \added{relationship between }luminosity \deleted{response}\added{and time} (normalized to a maximum value of 1) for a simple stellar population (SSP) in the left panel and for continuous star formation (CSF) in the right panel.  We find that for the CSF case, H$\alpha$ emission reaches [10\%, 50\%,90\%] of its maximal value after [1.32, 2.58, 5.03] Myr, while the NUV does so in [2.87, 24.56, 520.84] Myr.  These measurements provided by the newer stellar population synthesis models of FSPS provide updates to those from \cite{Kennicutt2012} and \cite{Hao2011}, who find that the corresponding values for the mean stellar ages contributing to emission for H$\alpha$ and the NUV are 3\,Myr and 10\,Myr respectively, while the stellar ages below which 90\% of emission is contributed are 10\,Myr and 200\,Myr respectively.  The discrepancy between timescales associated with 90\% of the maximum luminosity is largely due to the assumption of $Z = Z_\odot$ in \cite{Kennicutt2012} and \cite{Hao2011}, which differs from our assumption of $Z = 0.2\,Z_\odot$ here.

Having obtained the NUV and H$\alpha$ \deleted{luminosity responses}\added{time-dependent luminosities} from FSPS, we apply them to the simulated star formation histories by convolving the star formation histories with the SSP curves (normalized to have an integrated area of 1) shown in the left panel of Figure \ref{Fig: Response}.  Because we are given star formation histories for a set of simulated galaxies rather than a single observation for each, we can effectively increase the number of simulated measurements of which we can make use by utilizing every time step of each SFH and treating it as an individual galaxy measurement.  In this way, we use our 11,461 star formation histories to generate $\sim 10^8$ simulated measurements of the burst indicator.  

We plot the distribution of $\eta_\mathrm{th}$ and $\eta$ for the full SAM and MUFASA samples in Figure \ref{Fig: BSI}.  The suppression of the tail of the distribution for $\eta>0$ in the right panel of Figure \ref{Fig: BSI} is caused by the tendency for SFR$_\mathrm{NUV}$ to follow short timescale starbursts better than $\mathrm{\widetilde{SFR}}_{100}$.  The tail for $\eta < 0$ is unaffected because it is dominated by SFR$_\mathrm{H\alpha}$ going to 0 rather than the \added{time depedence of }NUV \deleted{luminosity response}\added{emission}.  The statistics for the distributions shown in Figure \ref{Fig: BSI} are listed in Table \ref{Tbl: simstats}.  It is important to note that $|\bar{\eta}| \leq 0.02$ for both the SAM and {\sc Mufasa} samples.  With the expectation that these cosmological simulations should be representative of observable galaxy populations, this implies that $\bar{\eta}$ is not a good tracer of galaxy star formation burstiness, in contrast with recent literature.  Rather, $\sigma_\eta$ provides a much more effective measure of burstiness.

\begin{figure*}
\centering
\includegraphics[width = \textwidth]{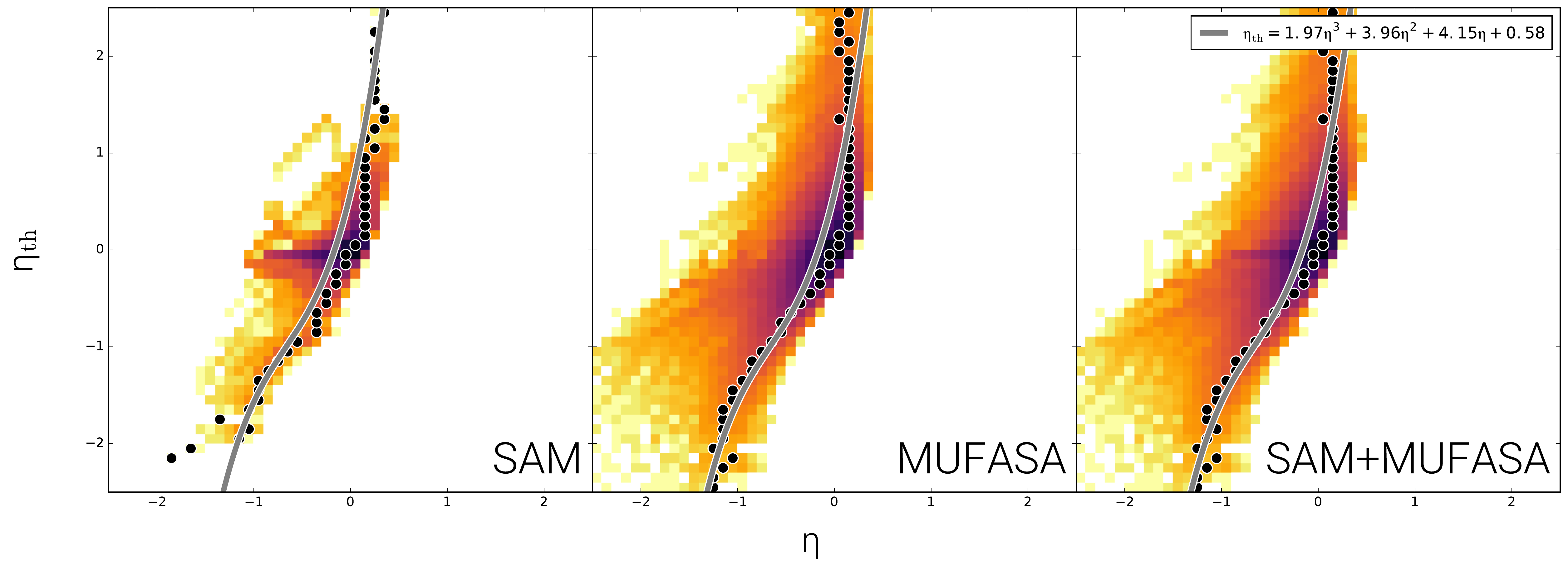}
\caption{\label{Fig: BSI Compare}A comparison between the observed and theoretical burst indicators, shown on the x- and y-axis respectively.  $\eta$ has been obtained by applying the \deleted{luminosity responses}\added{time-dependent luminosity curves} shown in Figure \ref{Fig: Response} and limiting to SFRs which would be observable using JWST+NIRSPEC.  Again, every time step for each galaxy of a given simulation is plotted, with colored squares indicating the density of points.  Darker regions indicate a higher density of points while lighter regions are less dense.  The black circles depict maximal density for each y-value.  The grey line is fit to these maxima in the right panel for both samples together and overplotted in the left and middle panels for comparison.}
\end{figure*}

With both the theoretical burst indicator ($\eta_\mathrm{th}$) and the \deleted{observed}\added{observational} burst indicator ($\eta$) in hand, we can now plot the relationship between the two so that a measurement of burstiness can potentially be converted into its theoretical counterpart.  We impose a JWST+NIRSPEC H$\alpha$ flux limit\footnote{The 5$\sigma$ flux limit we use for JWST+NIRSPEC is $2.0\times10^{-18}\,\mathrm{erg~s^{-1}}$ (see Section \ref{Sec: FluxLims}).} to our sample and plot the resulting relationship between $\eta$ and $\eta_\mathrm{th}$ for all remaining SFR measurements in Figure \ref{Fig: BSI Compare}.  There is a strong locus of points in the plot which we fit using ridge fitting.  The fit gives us the relationship between $\eta_\mathrm{th}$ and $\eta$, which we found was best modeled by cubic polynomial, as indicated in the figure legend.  These figures each show intervals where $\eta_\mathrm{th}$ and $\eta$ have a strong correlation ($-1 < \eta < 0$) and regions where $\eta$ is not as sensitive to changes in $\eta_\mathrm{th}$ because $\eta_\mathrm{th}$ is more sensitive to large changes in SFR over short timescales.  It should also be noted that $\eta$ represents an ideal \deleted{observed}\added{observational} estimator, which should only be applied to observed data after having corrected for dust attenuation.  It should be interpreted only after quantifying the effects of noise and other systematic effects in the data.

It is interesting to note that scatter about the ridge we fit is not uniform.  This asymmetric scatter is a product of the definition of the burst indicator combined with the properties of the simulated galaxies.  The simulations can have extended periods where the star formation rate is very close to zero, followed by a sudden upturn caused by the initiation of star formation.  Because the portion of $\eta_\mathrm{th}$ which represents continuous star formation is a 100\,Myr median, there are many instances where the numerator is a non-zero SFR value while the denominator is still approximately zero.  This generates very high measurements in comparison to $\eta$, which probes the continuous star formation rate via SFR$_\mathrm{NUV}$ which is analogous to a weighted average.  As a result of the comparatively greater sensitivity to sudden variations in star formation rate $\mathrm{\widetilde{SFR}}_\mathrm{100}$ has over SFR$_{NUV}$, $\eta_\mathrm{th}$ can be significantly larger than $\eta$ when galaxies transition away from a state of zero star formation.

% In addition, there is a horizontal line present in the left panel of Figure \ref{Fig: BSI Compare} at SFR$_{10}/$SFR$_{100}\approx 1.0$.  Upon further investigation, this appears to be a product of a specific feature which appears in the SAM SFHs where at late times they exponentially decline to a plateau before dropping to zero SFR.  This exponential decline and plateau creates a situation where SFR$_{10}/$SFR$_{100}\approx 1.0$, but SFR$_\mathrm{NUV}$ has not yet caught up to SFR$_\mathrm{H\alpha}$, causing the observed horizontal line.

\subsection{Population burstiness}

As was mentioned previously, it is observationally difficult for us to measure the burstiness of a single galaxy, although we have shown that it is possible to determine if it is in a period of recent rising, falling, or near-constant star formation rate.  It is, however, possible to characterize the burstiness of a population of galaxies from its distribution of burst indicators.  For example, the width of the distribution measures large deviations in star formation rate from a continuous rate, or a non-zero average $\eta$ could indicate an overall increasing or decreasing SFR for the sample.  The statistics of the entire sample of SFR measurements for each simulation represent the intrinsic burstiness of each and are summarized in Table \ref{Tbl: simstats}.

The average and skewness of $\eta_{th}$ for the SAM and {\sc Mufasa} simulations are both close to zero but there is some variation in the skewness.  This variation is caused by subtle qualitative differences between the SAM and {\sc Mufasa} simulations which are more difficult to detect in $\eta$.  The distributions of $\eta$ in the SAM and {\sc Mufasa} simulations both have average values of approximately zero, but they have differing standard deviations and are both negatively skewed.  The smoothed bursts of instantaneous star formation in the {\sc Mufasa} simulations (see Section \ref{Sec: Sims}) give us cause to expect that they would have a larger standard deviation because this effect will make them tend towards more \deleted{stochastic}\added{bursty} star formation histories than the SAMs.  This is seen as a larger measured $\sigma$ for MUFASA in both $\eta$ and $\eta_\mathrm{th}$ in Table \ref{Tbl: simstats}.  The negative skewness in $\eta$ is due to a tendency for galaxies to quench rapidly, dropping to effectively zero star formation ($\eta = -\infty$), while having no equivalent extreme for a starburst because a non-zero SFR$_\mathrm{H\alpha}$ necessitates a non-zero SFR$_\mathrm{NUV}$.  As a result, it is more likely to generate extreme low values on the plot than extreme high values.

\begin{figure*}
\centering
\includegraphics[width = \textwidth]{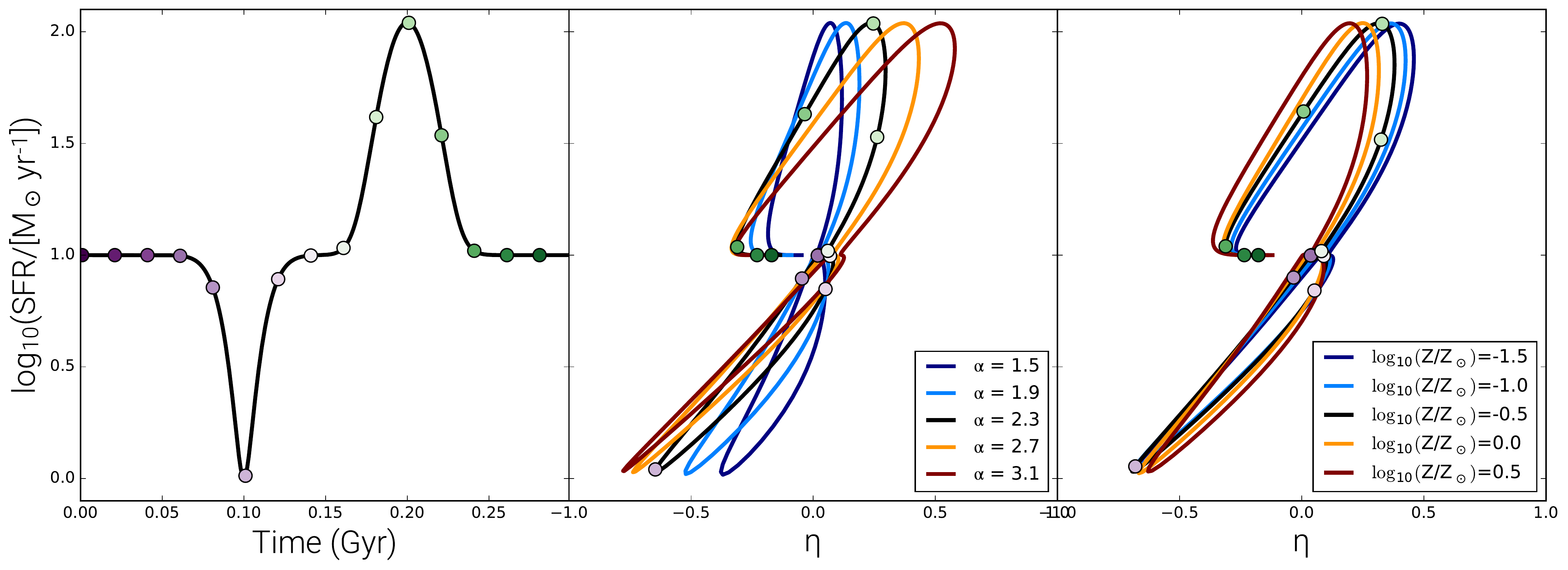}
\caption{\label{Fig: var metal IMF}This figure shows the change in burst indicator $\eta$ for a simple star formation history with a brief decrease and increase in SFR on 100\,Myr timescales. \textit{Left panel:} a synthetic SFH is plotted in black with colored points overplotted to easily compare early and late times to the right two panels.  The y-axis of the left panel is the instantaneous star formation rate while the y-axis of the middle and right panels is $\mathrm{SFR_{H\alpha}}$ (although they are nearly identical).  \textit{Middle panel:} shows the variation in $\eta$ for different values of high-mass IMF slope, as indicated in the legend.  The black line ($\alpha = 2.3$) matches slopes from \cite{Salpeter1959}, \cite{Chabrier2003}, and \cite{Kroupa2001}.  Steep IMF slopes can enhance the variance of $\eta$ while flatter slopes reduce it.  \textit{Right panel:} shows the variation in $\eta$ for different values of metallicity, indicated in the legend.  Variations in metallicity have very little effect on measurements of $\eta$.  Statistics for these plots are shown in Table \ref{Tbl: imf metal sfh}.}
\end{figure*}

\section{Effects of IMF, Metallicity, and Average SFH}\label{Sec: IMF, metal, SFH}

$\mathrm{SFR_{H\alpha}}$ and $\mathrm{SFR_{NUV}}$ (and therefore $\eta$) are sensitive to the high-mass slope of the IMF because O- and B-type stars dominate the H$\alpha$ and NUV emission respectively.  As a result, it is prudent to analyze the effects of changes in the IMF on star formation burstiness measurements.  Because burstiness can be estimated from the width of the $\eta$ distribution, we will pay particular attention to the effects various IMF slopes have on the standard deviation $\sigma$ of the distribution.  We start by recalculating the \added{relationships between time and H$\alpha$ and NUV emission shown in}\deleted{luminosity responses of} Figure \ref{Fig: Response} for IMF high mass slopes of $[1.5,~1.9,~2.3,~2.7,~3.1]$, where a high mass slope of $\alpha = 2.3$ agrees with \cite{Salpeter1959}, \cite{Chabrier2003}, and \cite{Kroupa2001}.  We then generate a test star formation history with a continuous star formation rate and a brief dip and burst in SFR, characterized by a gaussian with $\sigma = 12.5$\,Myr, to examine how these star formation histories affect $\eta$.

Our results are shown in the first two panels of Figure \ref{Fig: var metal IMF} and summarized in Table \ref{Tbl: imf metal sfh}.  The left panel of Figure \ref{Fig: var metal IMF} shows the mock star formation history while the middle panel shows the $\log_{10}\mathrm{SFR_{H\alpha}}$ vs. $\eta$.  Variation in the time average $|\bar{\eta}|$ with IMF slope is minimal ($<0.05$), even for this broad range of IMF slopes, however it is evident that the IMF can change the width of the distribution of measured $\eta$ values, with the range of standard deviations across these curves being $0.09 < \sigma < 0.28$ and $\sigma = 0.19$ for $\alpha = 2.3$ (see Table \ref{Tbl: imf metal sfh}).  The suppression of burstiness measurements for top-heavy IMFs emerges from the increased influence of O-type stars on the NUV luminosity and therefore SFR$_\mathrm{NUV}$.  This shortens the timescales probed by SFR$_\mathrm{NUV}$, bringing it closer to the timescales probed by SFR$_\mathrm{H\alpha}$.  Thus for a wide range of high-mass IMF slopes, the factor of suppression or enhancement of scatter in $\eta$ is $\lesssim2$.  As a result, variations in the IMF slope have little effect on the measured scatter of $\eta$.

Variations in metallicity are another factor which could potentially affect the measurement of $\eta$.  This result is shown in Figure \ref{Fig: var metal IMF} in the right panel.  We use the same synthetic SFH and plot SFR vs. $\eta$ for a range of metallicities.  It is evident that variations in metallicity have a significantly smaller effect on the distribution of $\eta$ here than in the case of variations in IMF.  The average $\eta$ is relatively unchanged by metallicity with $|\bar{\eta}| \leq 0.05$ for all metallicity values considered.  Standard deviations fall in the range $0.19 < \sigma < 0.123$ for the highest to lowest metallicity curves respectively, enhancing measurements of burstiness for sub-solar metallicities and suppressing it slightly for super-solar metallicities (see Table \ref{Tbl: imf metal sfh}).  As a result, metallicity has a negligible effect on measurements of $\eta$.

\begin{figure*}
\centering
\includegraphics[width = \textwidth]{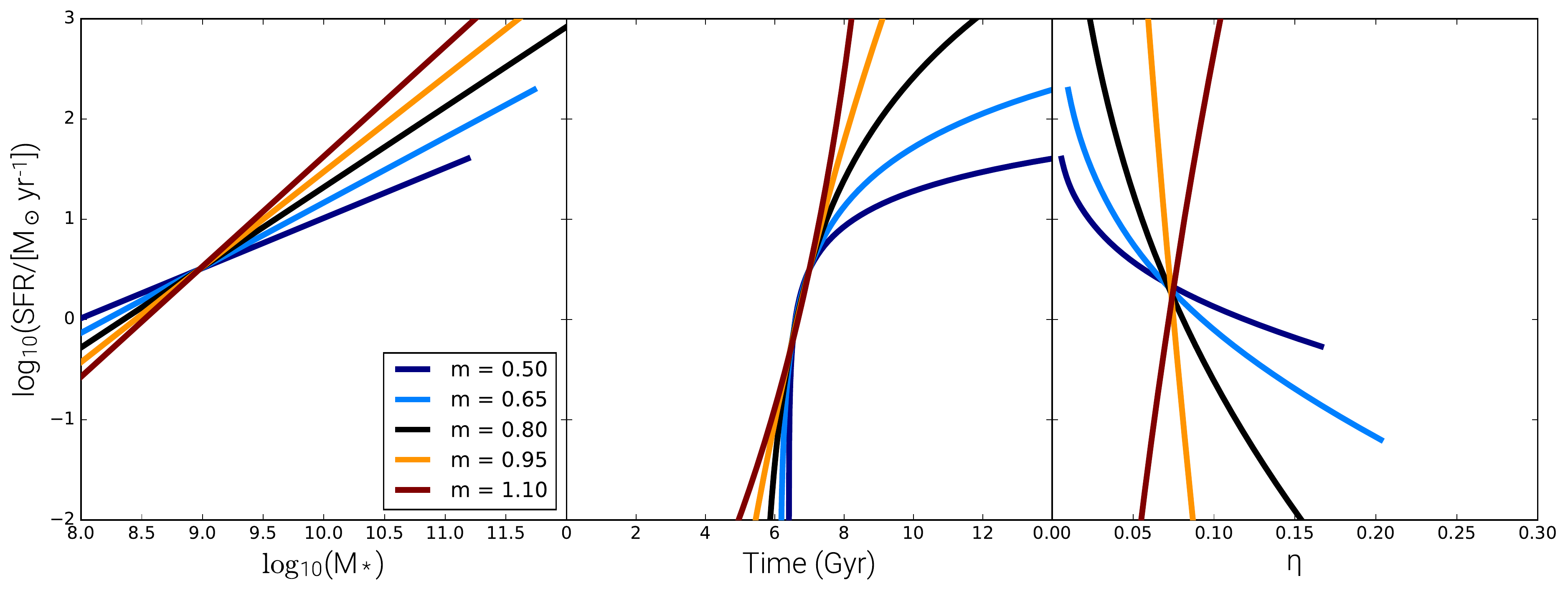}
\caption{\label{Fig: var SFH}A comparison of the effects of varying SFHs on the measurement of burst indicator $\eta$.  The left panel shows the power law SFHs which emerge under the assumption that galaxies evolve along an SFR-$M_*$ relation with constant slope and normalization.  Different values of SFR-$M_*$ slope are shown in the legend, with the black line representing the approximate expectation value of 0.8 for this redshift range.  Similar to Figure \ref{Fig: var metal IMF}, the y-axis of the left panel is the instantaneous SFR while the y-axis of the right panel is $\mathrm{SFR_{H\alpha}}$, which is nearly identical.  The right panel shows the corresponding measurements of $\eta$, indicating that depending on the mass (and therefore SFR) of a galaxy, shallower SFR-M$_*$ slopes cause greater deviations in $|\bar{\eta}|$ than steeper SFR-M$_*$ slopes.  Statistics for these plots are shown in Table \ref{Tbl: imf metal sfh}.}
\end{figure*}

In addition to variations in IMF, trends in average SFH for a population of galaxies such as a generally increasing or decreasing SFR across the sample will affect $\eta$ measurements.  To test this, we generate several mock SFHs which correspond to galaxy evolution along different slopes of the SFR-$M_*$ correlation.  This method corresponds to a simplified version of Main Sequence Integration, as described in \cite{Leitner2012}.  This is accomplished by starting from a reasonable midpoint SFR in logarithmic space($10^{0.5}~\mathrm{M_\odot~yr^{-1}}$) and finding its matching $\mathrm{M_*}$ from \cite{Kurczynski2016} which uses $\log_{10}(\mathrm{SFR/M_\odot\,yr^{-1}}) = b + m \log_{10}(\mathrm{M_*/M_\odot})$.  For $b=-6.903$ and $m=0.825$, this gives $\mathrm{M_*} = 9.4\times10^{8} \mathrm{M_\odot}$.  If we assume that each simulated SFH reaches this point at 7\,Gyr, then we can step time forwards and backwards from that point to get new SFR and $\mathrm{M_*}$ measurements at 1\,Myr intervals for the full range of time values ($0<t<14\,\mathrm{Gyr}$).

We then study how the measurement of $\eta$ would vary for a galaxy that evolve along an SFR-$M_*$ relation with a given slope and constant normalization, shown in Figure \ref{Fig: var SFH}.  The left panel shows the various SFHs which emerge for multiple values of the SFR-$\mathrm{M_*}$ correlation slope (indicated in the legend).  As we see in the right panel, flatter SFR-$\mathrm{M_*}$ slopes (shown in light and dark blue) can induce a modest change in the average measurement of $\eta$ at low SFR because these curves represent situations in which the stellar mass assembly is significantly faster than expected while at higher SFR, $\eta$ is close to zero.  Conversely, for the steepest slope (shown in red), early times and low SFRs do not significantly affect the average measurement of $\eta$.  At later times and higher SFRs however, the line moves away from $\eta = 0$.  A SFR-M$_*$ slope of $m=1$ corresponds to exponentially increasing SFHs and will cause no evolution in $\eta$.  We summarize the statistics in Table \ref{Tbl: imf metal sfh}, which shows that average SFH variations of this form cause minimal scatter in $\eta$ and a shift of $<0.1$ in its average value.

Finally, we consider several common SFHs used in the literature: constant, linearly rising, and exponentially declining with varying characteristic times \citep{Iyer2017}, with our results shown in Table \ref{Tbl: sfhstats}.  Constant star formation corresponds to no variation in $\eta$ from zero.  Linearly rising star formation measured at decreasing time intervals corresponds to increasing $\bar{\eta}$ and med($\eta$), but decreasingly positive $\sigma_\eta$ and skewness.  The exponentially declining SFHs have increasingly negative average and median values for $\eta$ with decreasing characteristic time ($\tau$).  The standard deviation increases positively as well for decreasing characteristic time while the skewness transitions from positive to negative.  We find that there is no significant difference in sensitivity to burstiness between $\bar{\eta}$ and similar values in the literature such as $\langle\mathrm{SFR_{NUV}/SFR_{H\alpha}}\rangle$.  To summarize, Table \ref{Tbl: sfhstats} shows that all of these simple parameterizations other than those which are strongly decreasing produce positive measurements of skewness.  Additionally, these SFHs typically have little measurable effect on $\bar{\eta}$, med($\eta$), or $\sigma_\eta$ except in the case where the exponentially declining SFH has $\tau$ significantly below 1\,Gyr or the linearly rising SFH is measured at $t\lesssim 300$\,Myr.  In this case, the SFH has structure on $\sim100$\,Myr timescales, which is detectable using $\eta$.
\added{We refer the reader to \cite{Meurer2009} Figure 8 and \cite{Iglesias2004} Table 3 for an earlier investigation of how burst indicators vary with SFH.}

\begin{deluxetable}{ccrcll}
\tablecaption{Statistics for IMF, Metallicity, and Average SFH Variations}
\vspace{-0.2cm}
\tablehead{\colhead{Property} & \colhead{Value} & \colhead{$\bar{\eta}$} & \colhead{$\mathrm{med(\eta)}$} & \colhead{$\sigma_\eta$} & \colhead{$\gamma_\eta$}}
\startdata
 & ~1.50 & -0.03 & ~0.00 & 0.09 & -1.19 \\
 & ~1.90 & -0.04 & ~0.01 & 0.14 & -0.96 \\
IMF Slope ($\alpha$) & ~2.30 & -0.04 & ~0.02 & 0.19 & -0.68 \\
 & ~2.70 & -0.03 & ~0.04 & 0.24 & -0.39 \\
 & ~3.10 & 0.01 & ~0.07 & 0.28 & -0.13 \\ \hline
 & -1.50 & 0.02 & ~0.08 & 0.23 & -0.42 \\
 & -1.00 & 0.00 & ~0.06 & 0.22 & -0.47 \\
$\log_{10}(Z/Z_\odot)$ & -0.50 & -0.02 & ~0.04 & 0.22 & -0.52 \\
 & ~0.00 & -0.04 & ~0.02 & 0.20 & -0.62 \\
 & ~0.50 & -0.05 & ~0.00 & 0.19 & -0.69 \\ \hline
 & ~0.50 & 0.03 & ~0.02 & 0.03 & ~2.65 \\
 & ~0.65 & 0.05 & ~0.03 & 0.04 & ~2.44 \\
SFH Slope (m) & ~0.80 & 0.07 & ~0.05 & 0.05 & ~2.24 \\
 & ~0.95 & 0.13 & ~0.11 & 0.07 & ~2.05 \\
 & ~1.10 & 0.12 & ~0.09 & 0.09 & ~3.71 \\
\enddata
\tablecomments{\label{Tbl: imf metal sfh}The average ($\bar{\eta}$), median, standard deviation ($\sigma_\eta$), and skewness ($\gamma_\eta$) are shown for the right two panels of Figure \ref{Fig: var metal IMF} and for Figure \ref{Fig: var SFH}.}
\end{deluxetable}

\begin{deluxetable*}{crcll}
\tablecaption{\deleted{Observed}\added{Observational} $\eta$ and Flux Ratios for Various SFHs}
\vspace{-0.2cm}
\tablehead{\colhead{SFH} & \colhead{$\bar{\eta}$} & \colhead{$\mathrm{med(\eta)}$} & \colhead{$\sigma_\eta$} & \colhead{$\gamma_\eta$}}
\startdata
Constant SFH & 0.00 & ~0.00 & 0.00 & ~0.00 \\
Linearly Rising ($t = 1\,$Gyr) & 0.14 & ~0.11 & 0.09 & ~3.02\\
Linearly Rising ($t = 300\,$Myr) & 0.24 & ~0.20 & 0.11 & ~2.31\\
Linearly Rising ($t = 100\,$Myr) & 0.36 & ~0.31 & 0.12 & ~1.74 \\
Exponentially Declining ($\tau = 3\,$Gyr) & -0.00 & -0.01 & 0.03 & ~1.52\\
Exponentially Declining ($\tau = 1$\,Gyr) & -0.06 & -0.07 & 0.06 & ~0.60\\
Exponentially Declining ($\tau = 300\,$Myr) & -0.75 & -0.60 & 0.62 & ~-0.51\\
\enddata
\tablecomments{\label{Tbl: sfhstats}The average ($\bar{\eta}$), median, standard deviation ($\sigma_\eta$), and skewness ($\gamma_\eta$) calculated for several common star formation histories (constant, linearly rising, and exponentially declining with various e-folding times).}
\end{deluxetable*}

\section{Mock Catalogs} \label{Sec: Mocks}

\begin{figure*}
\centering
\includegraphics[width = \textwidth]{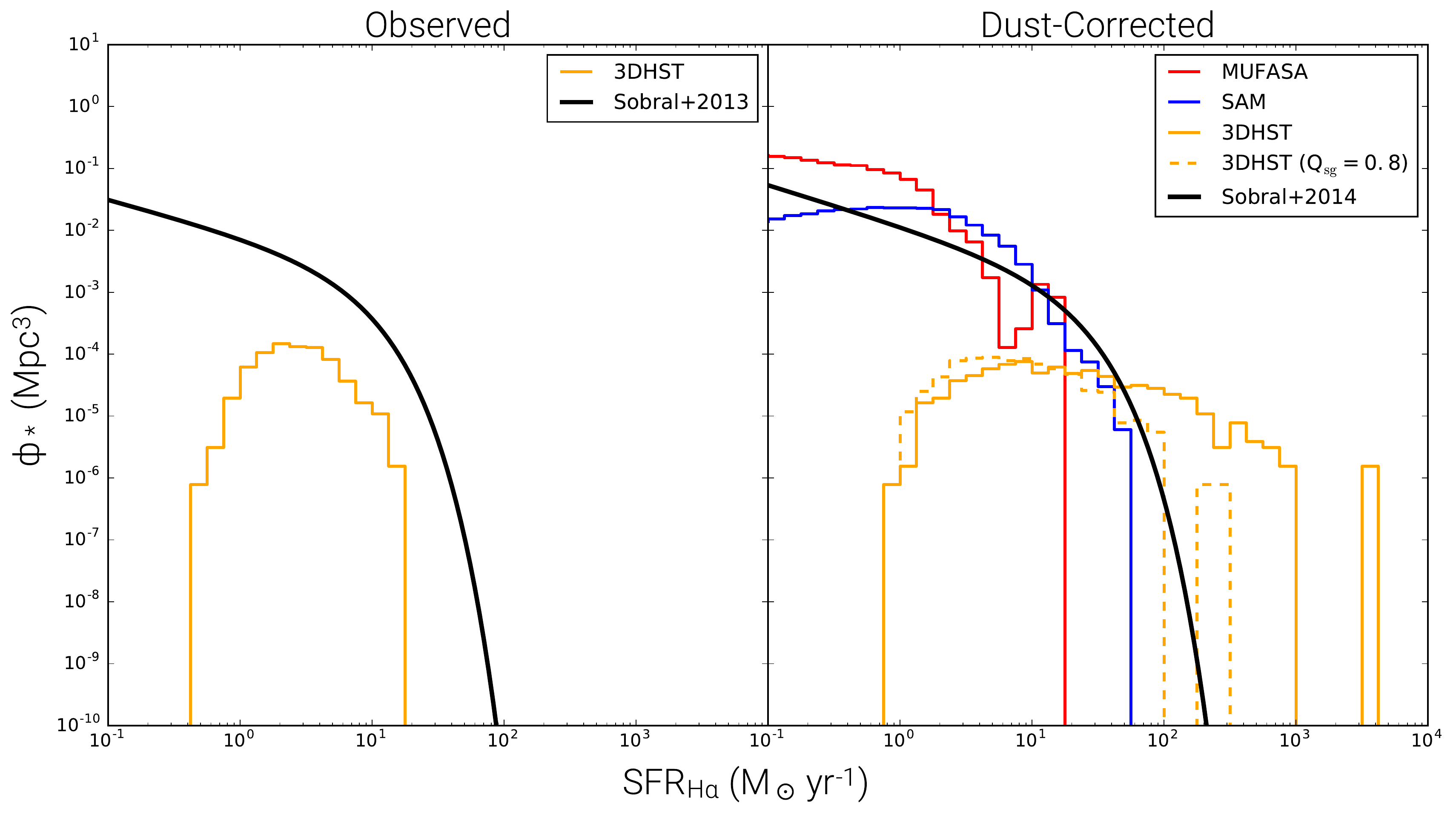}
\caption{\label{Fig: SFRF}The star formation rate functions (SFRFs) at $z\approx1$ for various samples used in this analysis.  The left panel shows the observed SFRF for the 3D-HST sample (orange line) and for data from \cite{Sobral2013}(black line).  The right panel shows the dust-corrected SFRFs for the 3D-HST and Sobral samples (shown as black and orange solid lines respectively) along with the SFRFs for the SAM and MUFASA simulated samples (blue and red lines respectively).  Finally, the orange dashed line represents the 3D-HST SFRF when using a dust prescription corresponding to $Q_\mathrm{sg}=0.8$ (see Section \ref{Sec: Gas, Star Dust}), and it agrees much better with the other samples.  It is evident that the standard dust-corrected 3D-HST sample contains galaxies which are significantly more luminous than any in the SAM, {\sc Mufasa}, and \cite{Sobral2014} \deleted{samples}\added{distributions}.  This could be evidence of dust overcorrection, possibly due to incorrect dust prescription, in the 3D-HST sample.  We examine the potential effects of incorrect measurement of dust reddening in Section \ref{Sec: Dust} and incorrect dust prescription in Section \ref{Sec: Gas, Star Dust}.}
\end{figure*}

In order to compare between the results found in the SAM and {\sc Mufasa} simulations and the 3D-HST data, we generate a mock catalog with the goal of matching the selection effects of the 3D-HST sample.  This catalog includes treatment of flux measurement errors, dust attenuation, and SFR measurement errors, along with the relevant applied flux limits.  In addition, galaxies are drawn from the simulations for the mock so that they match a star formation rate function (SFRF) from \cite{Sobral2014}.  However, when we plot the star formation rate functions (SFRFs) for the SAM and {\sc Mufasa} simulations and compare them with that of 3D-HST and \cite{Sobral2014}, some discrepancies arise (Figure \ref{Fig: SFRF}, right panel).  It seems that the 3D-HST sample contains $\mathrm{SFR_{H\alpha}}$ and H$\alpha$ luminosity measurements which reach $>1$ dex above those of the SAM or {\sc Mufasa} simulations or of \cite{Sobral2014}.  We have matched this sample to the observed sample from \cite{Pacifici2015} and found their measurements of the raw fluxes to be similar, indicating that this is not an issue in the raw data analysis.  We show in Section \ref{Sec: Dust} that $\eta$ is robust to the effects of incorrect measurement of dust attenuation under the assumption of a \cite{Calzetti2000} dust law.  We also find that this discrepancy largely disappears when plotting the same 3D-HST sample using a different dust prescription (right panel, orange dashed line), indicating that this could be an effect caused by an incorrect assumption of $E(B-V)_\mathrm{stars}/E(B-V)_\mathrm{gas}$ (see discussion in Section \ref{Sec: Gas, Star Dust}).  Using these analyses of dust measurement and dust prescription, we can make a useful comparison between the mocks and 3D-HST observed data.

Throughout this part of the analysis, each time bin measurement of a simulated galaxy's SFR is treated individually, meaning that single SFR measurements can be picked out to be placed in the mock catalog and that each drawn SFR is treated as a separate galaxy in the catalog.  

\begin{figure*}[t]
\centering
\includegraphics[width = \textwidth]{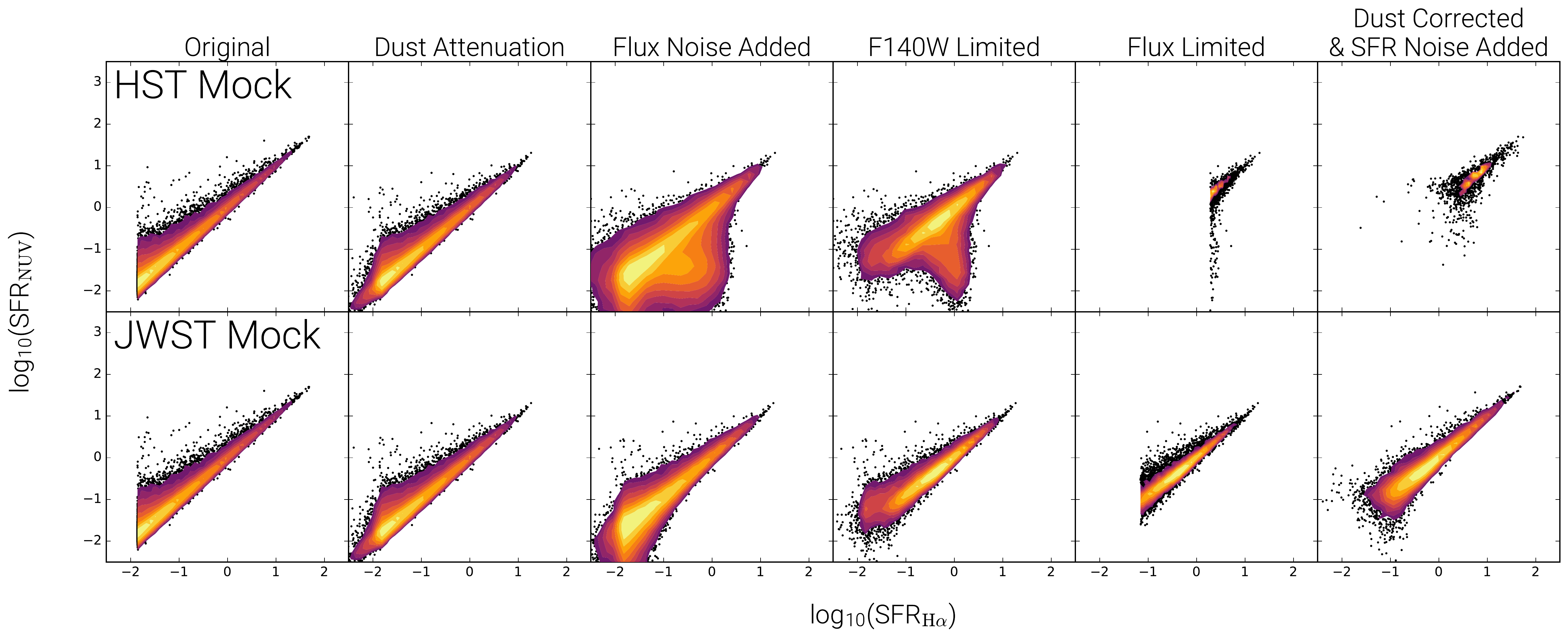}
\caption{\label{Fig: SFRProgression}Each panel is a plot of SFR$_\mathrm{NUV}$ vs. SFR$_\mathrm{H\alpha}$.  Each row shows how this distribution changes as selection effects are applied from left to right in the \added{3D-}HST mock (top) and the JWST mock (bottom).  We start with the original distribution drawn from the Schechter function in \cite{Sobral2014} (1st Column).  We then apply dust attenuation (2nd Column) and flux noise (3rd Column) and apply an F140W limit (4th Column) at the 5$\sigma$ detection limit of 25.8 magnitudes.  We further limit the sample based on H$\alpha$ flux detection limits (5th Column), correct for dust, and apply scatter based on the estimated SFR uncertainty (6th Column).}
\end{figure*}

The selection effects we apply here are shown in Figure \ref{Fig: SFRProgression}, which shows the progression in $\mathrm{SFR_{NUV}}$ vs. $\mathrm{SFR_{H\alpha}}$ as each selection effect is modeled.  We also generate a mock JWST catalog assuming usage of the NIRSpec instrument for H$\alpha$ measurement.  For clarity, throughout the remainder of this work we will refer to the addition of the effects of dust to simulated fluxes as dust attenuation while the removal of the effects of dust from simulated or observed fluxes will be referred to as dust correction.

\subsection{Star Formation Rate Function Selection}\label{Sec: SFRF}

To begin, we draw SFR measurements from the simulations in accordance with a SFRF based on observations.  We use the SFRF obtained by the High Redshift (Z) Emission Line survey (HIZELS; \citealt{Geach2008, Geach2009, Sobral2012, Sobral2013}) which is a ground-based deep narrowband H$\alpha$ survey in the range $0.4<z<2.2$.  The SFRF is reported in \cite{Sobral2014}; they fit a Schechter function to the observed distribution of SFR$_\mathrm{H\alpha}$ as follows:

\begin{align}\label{Eqn: Schechter}
\phi(SFR)~ d\,SFR= \phi^* \left( \frac{SFR}{SFR^*} \right)^\alpha e^{-SFR/SFR^*} d\left(\frac{SFR}{SFR^*}\right).
\end{align}
This fit is performed at various epochs for the dust-corrected $\mathrm{H}\alpha$ emission line.  Dust correction in \cite{Sobral2014} uses raw data from \cite{Sobral2013}, and it is based on the robust relationship between median stellar mass and median dust attenuation determined by \cite{Garn2010}.  This relationship was derived for a large sample of galaxies from the Sloan Digital Sky Survey (SDSS), but \cite{Sobral2012} have shown that this relationship holds to at least $z\sim1.5$.  3D-HST has an average redshift of $\bar{z}\approx 1$, which lies between fits at $z=0.84$ and $z=1.47$ in \cite{Sobral2014}, so we interpolate the fit parameters for these two redshifts to find a SFRF for $z\approx1$.  The resulting parameters are $\phi^* = 10^{-2.67}\,\mathrm{Mpc^{-3}}$, $SFR^* = 13.9\,\mathrm{M_\odot~yr^{-1}}$, and $\alpha = -1.66$.  We plot the observed SFRF for \cite{Sobral2013} and the 3D-HST sample along with the dust-corrected \cite{Sobral2014}, 3D-HST, SAM, and {\sc Mufasa} data in Figure \ref{Fig: SFRF}.

We normalize this curve to $1$ by integrating over the range $10^{-3} < SFR/SFR^* < 10^2$, and we use the resulting normalization factor $N$ to convert the SFRF into a probability distribution function (PDF), $p(SFR) \equiv \phi(SFR)/N$.  We then draw $10^5$ SFRs from this distribution.  A matching SFR within 0.5 dex of the drawn SFR is randomly chosen from the SAMs or {\sc Mufasa} in the epoch $0.8<z<1.2$.  Then $\mathrm{SFR_{H\alpha}}$, $\mathrm{SFR_{NUV}}$, and $M_*$ for the chosen SFH are rescaled so that the galaxy's $\mathrm{SFR_{H\alpha}}$ exactly matches the one drawn from the PDF.

We choose a ratio of SAM galaxies drawn to {\sc Mufasa} galaxies drawn for the mocks that leaves an approximately equal number of SAM and {\sc Mufasa} galaxies after the remainder of the selection effects in this section have been applied to the data for the 3D-HST depth case.  We are left with a starting mock catalog that consists of 1026 star formation rates, chosen near $z=1$ with a distribution of SFRs that matches the SFRF interpolated from \cite{Sobral2014}.

\subsection{Dust Attenuation}\label{Sec: Dust}

\begin{figure}
\centering
\includegraphics[width = 0.42\textwidth]{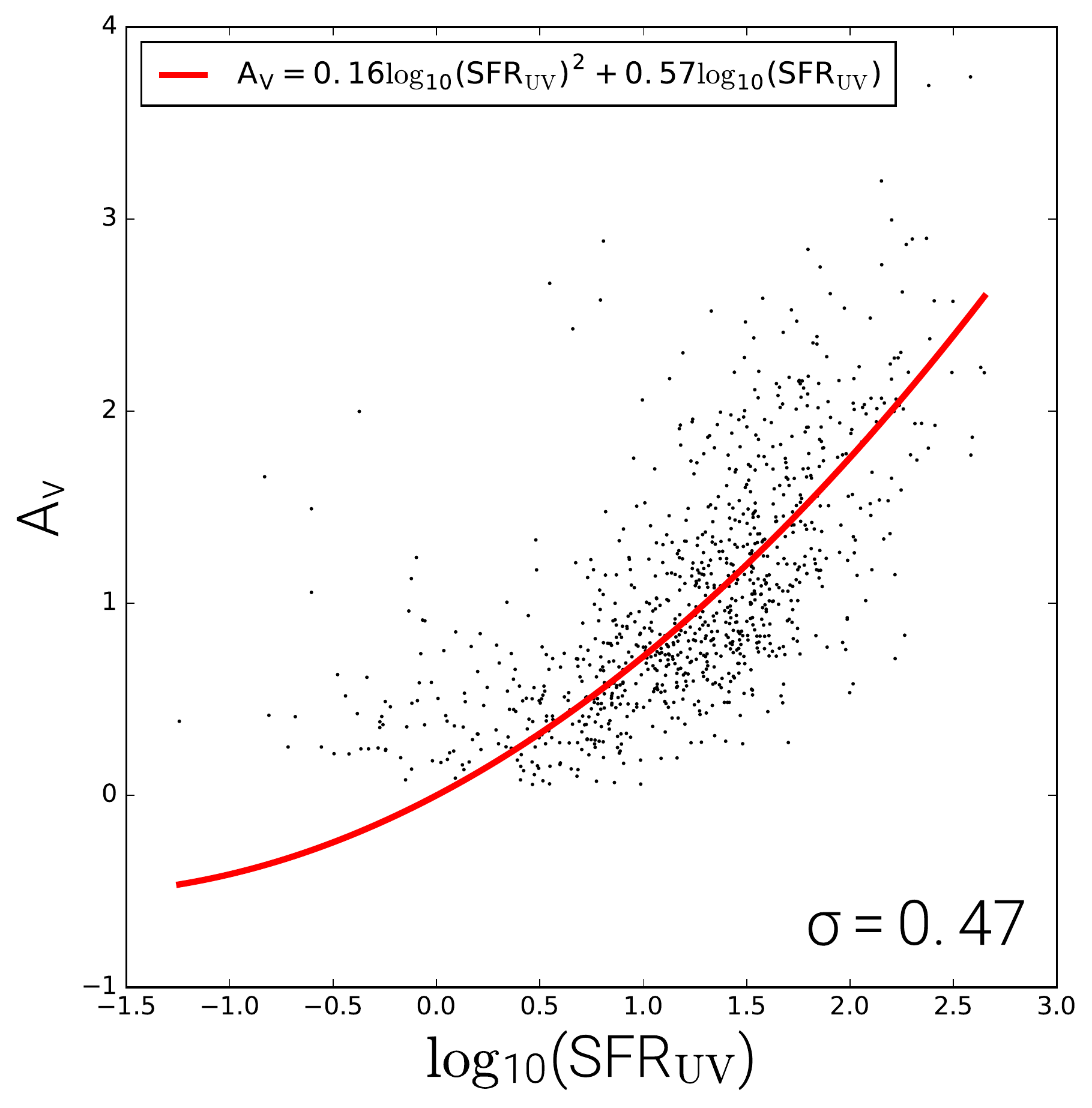}
\caption{\label{Fig: SFR100Av}The relationship between $A_V$ and $\mathrm{SFR_{100}}$ for 3D-HST galaxies.  Here, $\sigma$ represents the scatter of the 3D-HST data about the red fit line.  The fit and scatter shown are used in Section \ref{Sec: Dust} to assign values of $A_V$ to galaxies in the mock catalog.}
\end{figure}

After drawing the mock catalog, we convert the star formation rates to fluxes in order to apply dust attenuation.  First, we fit the measured $A_V$ values for the 3D-HST sample as a function of the ultraviolet SFR derived from SED fitting, as shown in Figure \ref{Fig: SFR100Av}.  This relationship is preferred over a fit to $A_V$ vs. $\mathrm{SFR_{H\alpha}}$ because SFR$_\mathrm{UV}$ does not have the strict selection effect we find in SFR$_\mathrm{H\alpha}$ at fixed redshift in 3D-HST, allowing us to better estimate the scatter in $A_V$.   Assuming Gaussian scatter, we assign $A_V$ values to the mock catalog galaxies based on their measured SFR$_\mathrm{NUV}$.  Finally, we make use of the dust law from \cite{Calzetti2000} to attenuate the NUV and H$\alpha$ flux under the assumption that $E(B-V)_{\mathrm{NUV}} = 0.44 E(B-V)_{\mathrm{H\alpha}}$ \citep{Calzetti1997,Wuyts2011} and $R'_v$ = 4.05 \citep{Calzetti2000}.

One of the primary conclusions in \cite{Guo2016} was that improper measurement of dust attenuation has very little effect on the ratio of far ultraviolet flux density to H$\beta$ flux.  Under the same \cite{Calzetti2000} dust law, they find $k(4861\mathrm{\,\AA})/0.44 - k(1500\,\mathrm{\AA}) \sim 0.1$.  We find a similar relationship here using H$\alpha$ and the NUV, which gives $k(6563\mathrm{\,\AA})/0.44 - k(2750\mathrm{\,\AA}) \sim 0.2$.  

We can derive the dependence of $\eta$ on the dust reddening:

\begin{align}\label{Eqn: Eta Dust}
\eta & = \log_{10}\left( \frac{C_1 F_\mathrm{H\alpha,obs}10^{0.4k_\mathrm{H\alpha}E(B-V)_\mathrm{stars}/0.44}}{C_2 F_\mathrm{\nu,NUV,obs}10^{0.4k_\mathrm{NUV}E(B-V)_\mathrm{stars}}} \right) \nonumber\\
\eta & = \log_{10}\left( \frac{C_1 F_\mathrm{H\alpha,obs}}{C_2 F_\mathrm{\nu,NUV,obs}} \right) + \nonumber \\ & \hspace{0.8cm} 0.4\left[ \frac{k_\mathrm{H\alpha}}{0.44} - k_\mathrm{NUV} \right]E(B-V)_\mathrm{stars}.
\end{align}
Here, $C_1$ and $C_2$ represent constants involved in the conversion from H$\alpha$ and NUV star formation rates to their respective flux densities, $F_\mathrm{H\alpha}$ and $F_\mathrm{NUV}$.  We then find the dependence of $\eta$ on the error in our measurement of dust reddening from the derivative with respect to $E(B-V)_\mathrm{stars}$.

\begin{align}
\frac{d\eta}{dE(B-V)_\mathrm{stars}} & = 0.4\left[ \frac{k_\mathrm{H\alpha}}{0.44} - k_\mathrm{NUV} \right] \approx -0.08
\end{align}
For comparison, a similar analysis using the H$\beta$ emission line and FUV broadband photometry yields $d\eta/dE(B-V)_\mathrm{stars} \approx 0.04$.

As an extreme example, consider a galaxy for which the NUV luminosity is overestimated by a factor of 100 i.e., a 5 magnitude error.  For this example case, we have $k_\mathrm{NUV}\approx7.35$ which gives an $E(B-V)_\mathrm{stars}$ error of $\sim 0.68$ and therefore an error in $\eta$ of $\sim -0.054$.  As a result, we find that $\eta$ is sensitive only to extreme errors in dust measurement.  As we will see in Section \ref{Sec: Gas, Star Dust}, independence between $\eta$ and dust measurement relies on the assumption that the correct value of $E(B-V)_{\mathrm{stars}}/E(B-V)_{\mathrm{gas}}$ has been used in the dust correction.

\subsection{Flux Uncertainties}\label{Sec: Fluxerr}

We approximate the uncertainty of each H$\alpha$ flux measurement in the \added{3D-}HST mock catalog by first fitting the relationship between measured H$\alpha$ flux and H$\alpha$ flux uncertainty in the 3D-HST sample.  This yields $\sigma_{F_\mathrm{H\alpha}} = \sqrt{1.25\times10^{-18}~\mathrm{erg~s^{-1}~cm^{-2}}~ F_\mathrm{H\alpha}}$ with a gaussian scatter of $9.98 \times 10^{-18}~\mathrm{erg~s^{-1}~cm^{-2}}$, which we apply to obtain the H$\alpha$ flux uncertainties.  In the case of the JWST mock, the flux uncertainty used is equivalent to the 3D-HST flux uncertainties without the gaussian scatter multiplied by a factor of $\sigma_{F_\mathrm{JWST}}/\sigma_{F_\mathrm{3D-HST}}$, which is an improvement by a factor of $~28$ (see Section \ref{Sec: FluxLims}).  In order to approximate the NUV uncertainty, we utilize the flux uncertainty vs. flux relationship for the $F606W$ band in the GOODS-S wide field survey.  These values follow the relationship $\sigma_{F_\mathrm{\nu,NUV}} = \sqrt{1.89\times10^{-32}\,\mathrm{erg\,s^{-1}\,cm^{-2}\,Hz^{-1}}~ F_\mathrm{\nu,NUV}}$.  Gaussian random numbers are drawn, with these uncertainties describing the standard deviation, in order to add noise to the H$\alpha$ (NUV) fluxes (flux densities).

\subsection{F140W Detection}

In order for a given galaxy to be detected in H$\alpha$ in 3D-HST, it requires a detection in the $F140W$ filter, which acts as a calibration image for the grism data.  By analyzing expected F140W fluxes for a grid of stellar mass and star formation rate values, we found that a flux limit at 25.8 AB magnitudes is represented in SFR-$M_*$ space by the following expression:
\begin{align*}
& \left[\log_{10}M* > 8.773\right] ~\vee~ \left[\log_{10}\mathrm{SFR_{H\alpha}} > -0.123\right] \\& \hspace{0.4cm} \vee~ \left[\left( \log_{10}M_* - (8.773-2) \right)^2 \right. \times \\ &\hspace{0.8cm} \left. \left( \log_{10}\mathrm{SFR_{H\alpha}} - (-0.123 - 2) \right)^2 > 6\right].
\end{align*}
Galaxies outside this region would not be present in 3D-HST.  We assumed the same continuum detection limit for our JWST mocks as well.

\subsection{H$\alpha$ Flux Limit}\label{Sec: FluxLims}

\cite{Brammer2012a} notes that the $5\sigma$ detection limit of the G141 grism is $F_\mathrm{H\alpha} = 5.5\times 10^{-17}~\mathrm{erg~s^{-1}}$, and as a result, we apply this as a flux limit on the mock catalog.  The $5\sigma$ detection limit of the Near Infrared Spectrograph (NIRSpec) is used for the JWST mocks and is approximately $2.0\times10^{-18}~\mathrm{erg~s^{-1}}$ for a 1-hour exposure of H$\alpha$ at $z\approx1$\footnote{From the JWST Pocket Guide: https://jwst.stsci.edu/files/live/sites/jwst/files/home/\\instrumentation/technical\%20documents/jwst-pocket-guide.pdf}.  This is an improvement of a factor of $\sim28$ over the G141 sensitivity employed by 3D-HST.

\subsection{Dust Correction \& SFR Noise}\label{Sec: FinalCut}

Finally, we correct for dust in the mocks and apply an SFR measurement scatter to $\mathrm{SFR_{NUV}}$ and $\mathrm{SFR_{H\alpha}}$ as well, using the 3D-HST-measured relationship between the SFR uncertainty and the H$\alpha$ SNR.  We find the correlation between the two to be $\sigma_\mathrm{SFR_{H\alpha}} = 1.368 (F_\mathrm{H\alpha}/\sigma_{F_\mathrm{H\alpha}})^{-0.066}$ and $\sigma_\mathrm{SFR_\mathrm{NUV}} = 1.832 (F_\mathrm{H\alpha}/\sigma_{F_\mathrm{H\alpha}})^{-0.102}$ for the H$\alpha$ and NUV, respectively.  This method assumes a fixed fractional uncertainty in SFR$_\mathrm{H\alpha}$ for values with the same SNR.  After having implemented uncertainties, noise, and flux limits to the mock catalogs, we can return to the starting measurements of $\eta$ (before adding noise) for the selected galaxies and find that the range of true $\eta$ values for our 3D-HST and JWST mock catalogs are $-0.47 < \eta < 0.27$ and $-1.24 < \eta < 0.23$ respectively. 

\begin{figure*}
\centering
\includegraphics[width = \textwidth]{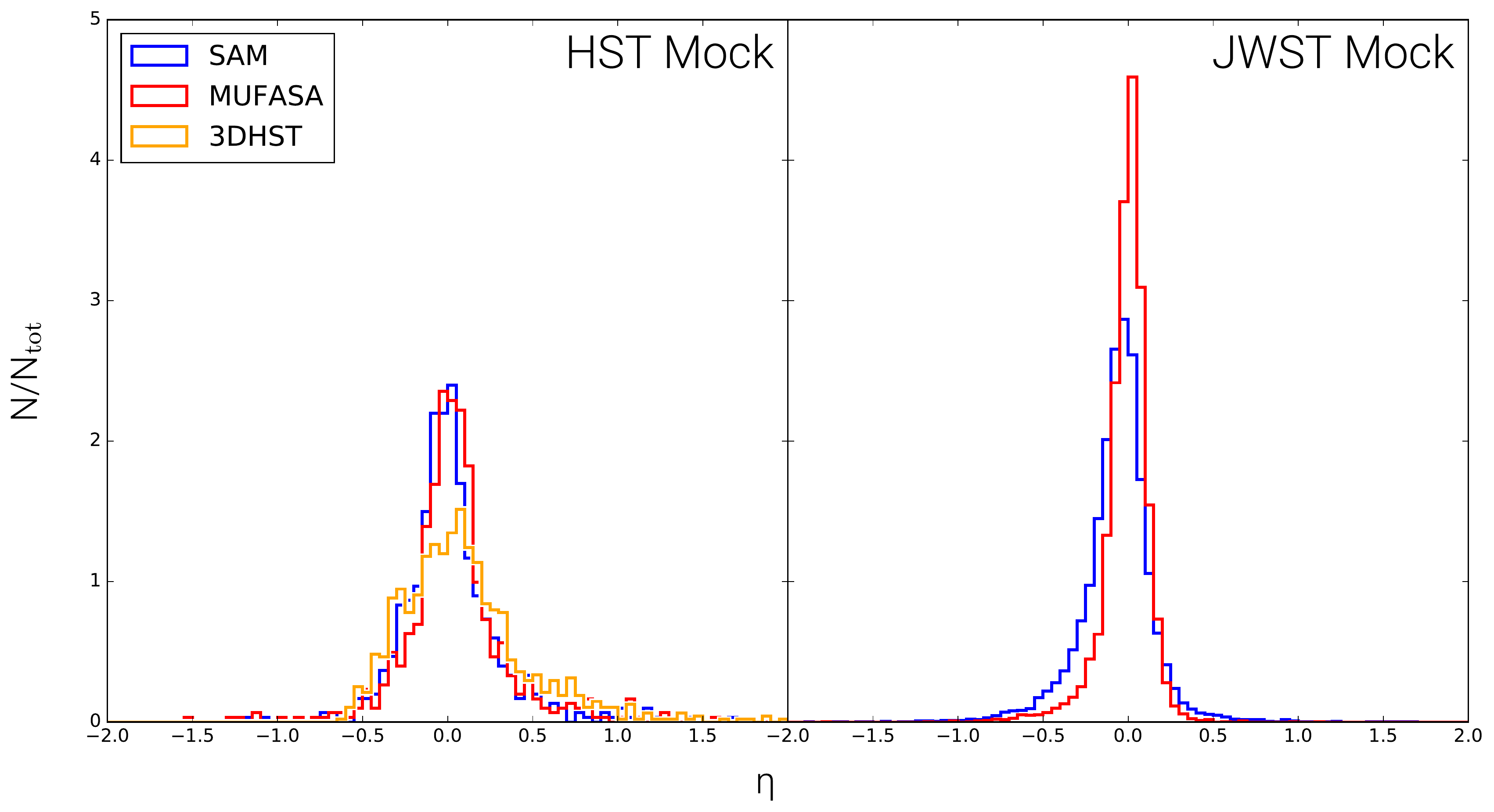}
\caption{\label{Fig: Burstiness Hist Both}The normalized burst indicator ($\eta$) distribution for the \added{3D-}HST mocks and 3D-HST data (left panel) and the JWST mocks (right panel), separated by simulation.  The SAM galaxies in each mock are represented by the blue histogram while the {\sc Mufasa} galaxies are represented in red.  The 3D-HST data are shown in orange\added{ for comparison} with the \added{3D-}HST mocks\deleted{ for comparison}.  The distribution of the JWST mocks is significantly \deleted{more }narrow\added{er} than that of the \added{3D-}HST mocks, largely due to a significant leap in measurement accuracy.  The JWST mocks indicate that a \deleted{very accurate}\added{more precise} measurement of burstiness\deleted{ and comparison between the simulations and observed data} will be possible with the incorporation of JWST data.  Statistics for these plots are given in Table \ref{Tbl: mockstats}.}
\end{figure*}

\section{Results}\label{Sec: Results}

\deleted{With our mock catalog generated from the SAM and {\sc Mufasa} simulated galaxies, and the systematics of the 3D-HST sample implemented, w}\added{W}e can now plot the burstiness of the mock \added{3D-HST observed }catalog and compare it to that of the \added{observed }3D-HST sample (Figure \ref{Fig: Burstiness Hist Both}; left panel).  We find moderate disagreements in the distribution width between the 3D-HST sample and the SAM and {\sc Mufasa} subsamples \deleted{within} the 3D-HST mocks.  This could be the result of genuinely greater levels of \deleted{star formation stochasticity}\added{burstiness} in the 3D-HST sample than the simulations predict, or \deleted{that there is}\added{of} an additional source of scatter which has not been taken into account.  One potential source of systematic error in burstiness measurements quoted in other analyses is stochastic sampling of the IMF, or in general, variations in IMF at low SFR.  However, our analysis is observationally limited to $\mathrm{SFR}\gtrsim1\,\mathrm{M_\odot\,yr^{-1}}$,  which is well in excess of the low SFRs where stochastic sampling of the IMF is significant ($\mathrm{SFR}\gg 0.01\,\mathrm{M_\odot\,yr^{-1}}$; \citealt{daSilva2012}).  \deleted{In an effort to ensure that}\added{As a cross check upon} the SpeedyMC SED fits we use for the 3D-HST data\added{,}\deleted{ do not greatly differ from}\added{ we performed the same analysis using dust measured by} the FAST \citep{Kriek2009} fits reported by \cite{Skelton2014}\deleted{, we performed the same analysis using FAST-measured dust}, finding no significant changes in the distribution of $\eta$.

The right panel of Figure \ref{Fig: Burstiness Hist Both} shows the same distribution for the mocks with \deleted{estimated} JWST selection effects applied\added{.  JWST}\deleted{, which} will provide a significant improvement to our measurements of SFR$_\mathrm{H\alpha}$ over the current 3D-HST data.  As is evident in the \deleted{rightmost two panels}\added{transition from panel 5 to panel 6 in the top row} of Figure \ref{Fig: SFRProgression}, there is a significant amount of scatter which comes into the distribution in the \deleted{final} step\deleted{, which is adding} \added{that adds }scatter based on SFR \added{measurement }uncertainties.  This is \deleted{particularly} evident when we compare the burst indicator distribution between the \added{3D-}HST mocks and the JWST mocks (Figure \ref{Fig: Burstiness Hist Both}).  The distribution of $\eta$ values of the 3D-HST mocks is \deleted{much} wider than that of the JWST mocks due to the dominant effects of scatter caused by the 3D-HST uncertainties, which will be greatly reduced by JWST.  We summarize the statistics of Figure \ref{Fig: Burstiness Hist Both} in Table \ref{Tbl: mockstats}.  It is evident by comparing $\sigma_\eta$ for the 3D-HST and JWST mocks against $\sigma$ for the SAM and {\sc Mufasa} simulations in Table \ref{Tbl: simstats} that the JWST mocks are able to represent the burstiness inherent in the simulations much more accurately than the 3D-HST mocks\deleted{ because the JWST mocks are not dominated by systematic scatter and represent a significantly larger population}.

\begin{figure*}
\centering
\includegraphics[width=\textwidth]{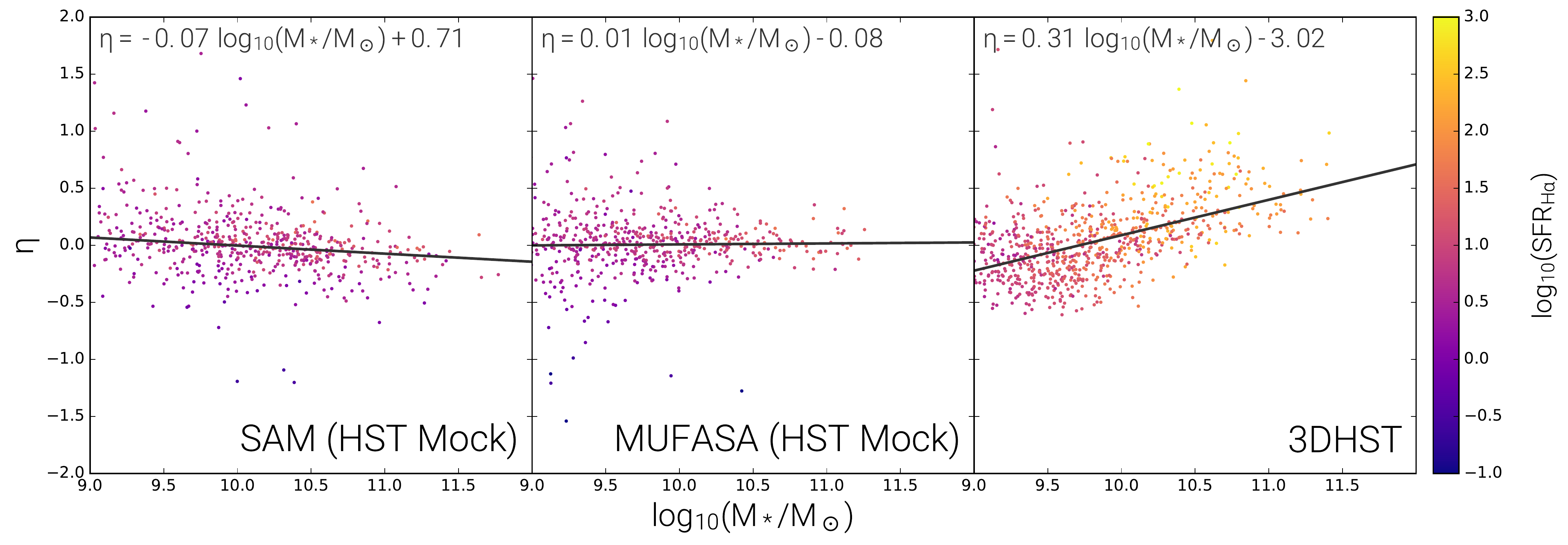}
\caption{\label{Fig: Burstiness HST}The distribution of $\eta$ as a function of stellar mass for the SAM galaxies (left) and {\sc Mufasa} galaxies (middle) in the \added{3D-}HST mock catalog, along with the 3D-HST observed data (right).  Points are colored according to their H$\alpha$ star formation rate.  From the grey fit lines plotted in each panel, it is evident that measurements of $\eta$ in the mock catalogs have minimal dependence on stellar mass, while the 3D-HST sample shows a positive trend.}
\end{figure*}

We also show the relationship between $\eta$ and $M_*$ for the 3D-HST sample (Figure \ref{Fig: Burstiness HST}) and find that there is a positive correlation between $\eta$ and $M_*$ for the 3D-HST sample \deleted{only}\added{that is not present in the mock sample}.  This trend could indicate either a genuine dependence in $\eta$ on $M_*$ or a systematic effect present in the 3D-HST sample that has not been implemented in the mocks.  We \deleted{performed}\added{repeated} the same analysis using FAST-measured dust here as well, again finding that there was no significant change from the $\eta$-$M_*$ trend we observed using the SpeedyMC fits.  In Section \ref{Sec: Gas, Star Dust}, we show that changing the dust prescription has the potential to partially resolve this discrepancy.

\begin{figure*}
\centering
\includegraphics[width = \textwidth]{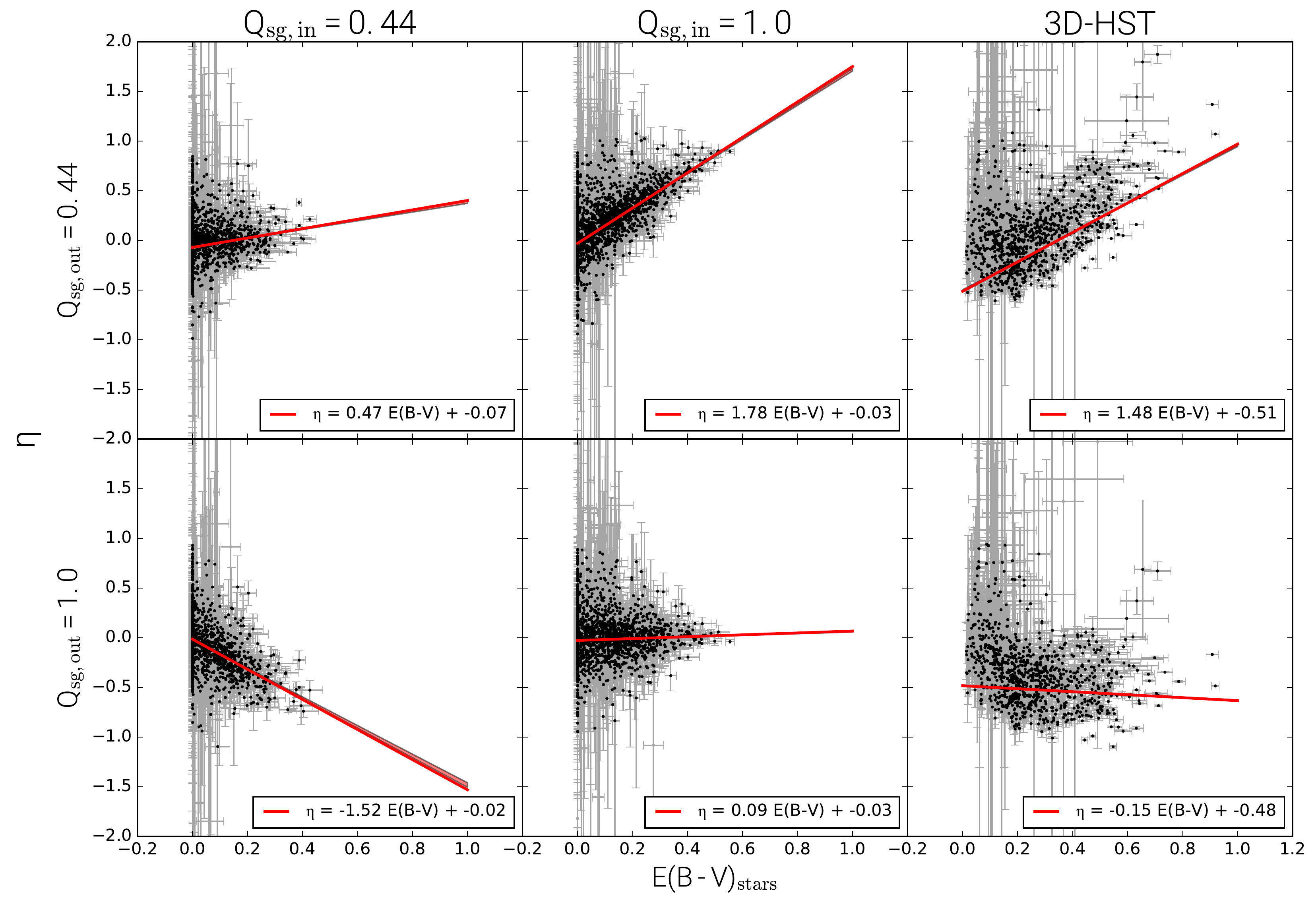}
\caption{\label{Fig: eta dust sigmaclip}Plot of $\eta$ vs. $E(B-V)_\mathrm{stars}$ for various input and output values of $Q_\mathrm{sg}$, as indicated by the column and row titles.  \added{The first two columns of panels represent the 3D-HST mocks, while the third column represents the observed 3D-HST data.  }The red line indicates the best fit to the data after performing iterative $3\sigma$ sigma clipping to the sample.  The black points indicate points which are not considered outliers by the fitting algorithm, while grey points are outliers with error bars shown in grey as well.  The red shaded regions represent the regions between the 16th and 84th percentile fits to the data after 1000 Monte Carlo realizations.  The broader observed range of $E(B-V)_\mathrm{stars}$ in the middle column (compared to the left) is an effect of $Q_\mathrm{sg,in} = 1.0$ causing dust to be less opaque for H$\alpha$ emission, allowing more galaxies above the H$\alpha$ limit imposed in Section \ref{Sec: FluxLims} for a given value of $E(B-V)_\mathrm{stars}$.}
\end{figure*}

\begin{figure}
\centering
\includegraphics[width = 0.5\textwidth]{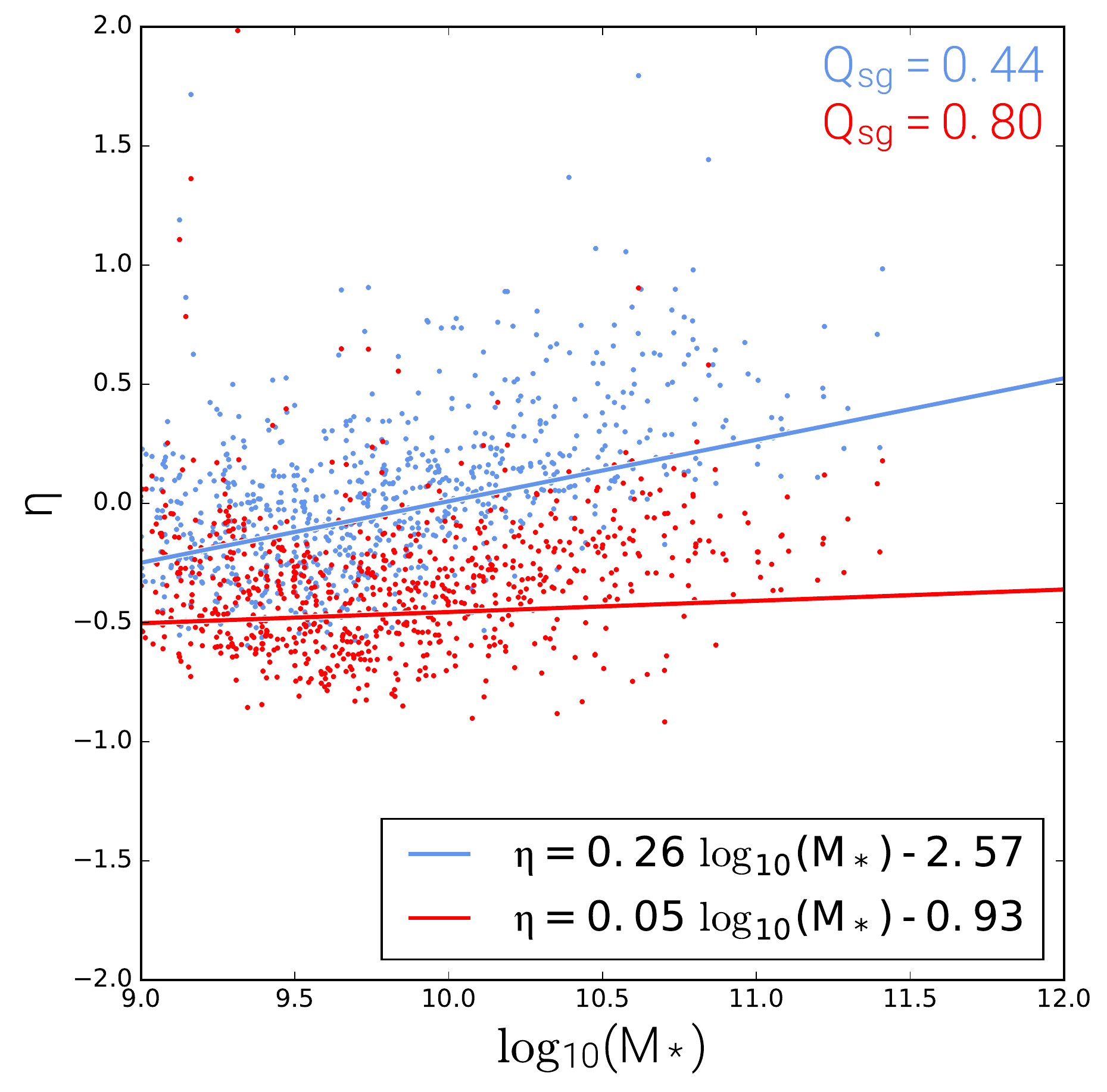}
\caption{\label{Fig: eta mstar}A plot of $\eta$ vs $\mathrm{M_*}$ for the 3D-HST data, similar to Figure \ref{Fig: Burstiness HST} right panel.  The light blue points and fit assume $Q_{sg} = 0.44$ in accordance with literature.  Red points and fit have been recalculated for $Q_{sg} = 0.8$, which is the value that gives linear independence between $\eta$ and $E(B-V)_\mathrm{stars}$ (see Section \ref{Sec: Gas, Star Dust}).}
\end{figure}

\begin{deluxetable}{cclll}
\tablecaption{$\eta$ Distribution Statistics for Mock Catalogs and 3D-HST}
\vspace{-0.2cm}
\tablehead{  &  & \colhead{$\bar{\eta}$} & \colhead{$\sigma_\eta$} & \colhead{$\gamma_\eta$}}
\startdata
3D-HST Data &  & \phantom{-}0.119 & 0.495 & \phantom{-}5.355\\ \hline
3D-HST Mock & {\sc Mufasa} & \phantom{-}0.044 & 0.317 & \phantom{-}0.402 \\
 & SAM & \phantom{-}0.024 & 0.302 & \phantom{-}1.119 \\ \hline
JWST Mock & {\sc Mufasa} & -0.013 & 0.142 & -1.724 \\
 & SAM & -0.066 & 0.225 & -0.216 \\
\enddata
\tablecomments{\label{Tbl: mockstats}Here we show the average ($\bar{\eta}$), standard deviation ($\sigma$), and skewness ($\gamma$) for the portions of galaxies in each mock which are drawn from SAM or {\sc Mufasa}.}
\end{deluxetable}

\section{Discussion}\label{Sec: Gas, Star Dust}

In section \ref{Sec: Dust}, we found that measurements of $\eta$ are robust to the effects of incorrect measurements of dust attenuation\deleted{, h}\added{  H}owever\added{,} this was under the assumption of the correct ratio of $E(B-V)_{\mathrm{stars}}$ to $E(B-V)_{\mathrm{gas}}$, which we will call $Q_\mathrm{sg}$.  \cite{Calzetti1997} finds that $Q_\mathrm{sg}=0.44$ by utilizing the Balmer decrement and assuming a `standard' extinction curve, and this result is confirmed by \cite{Wuyts2011}.  This is believed to hold for low redshifts, however \added{there is some disagreement at $z>1$.  For example, }\cite{Reddy2010} have indicated that $Q_\mathrm{sg} = 1.0$ is more accurate at $z\sim2$\added{, while \cite{Price2014} finds that $Q_{sg} = 0.5\pm0.3$ at $z\sim1.5$}.  In this analysis, we will assume that $Q_\mathrm{sg}$ is a constant across all galaxies in a given sample, however, it is important to keep in mind that measured dust prescriptions can vary greatly based on dust geometry alone \citep{Narayanan2018, Witt2000, Gordon1997, Calzetti1994}.

We compare these two scenarios for the mock catalogs and for the 3D-HST sample, shown in Figure \ref{Fig: eta dust sigmaclip}\deleted{ as each row}, by plotting the relationship between $\eta$ and $E(B-V)_\mathrm{stars}$ in each case.  Because we are able to apply dust attenuation to the mock catalogs in any way we choose, we can vary whether dust attenuation is inserted into the mock catalogs using $Q_\mathrm{sg}=0.44$ or 1.0 as well as \deleted{how its effects are corrected}\added{which value is assumed when correcting for dust}.\deleted{  This effect is shown in the figure, as each row is labeled with the factor used to correct mock observations for dust, while the first two columns are labeled with the factor used when modeling dust attenuation for each mock galaxy.  The third column shows the actual 3D-HST data, which are dust-corrected using the given value of $Q_\mathrm{sg}$.}  An error-weighted fit is performed on all of the data; points which are more than $3\sigma$ from the fitted line are removed from the set of points used for fitting, and the process is repeated until no more points are removed.  This gives a fit which is robust to outliers. 

We can derive the expected relationship between $\eta$ and $E(B-V)_\mathrm{stars}$ for any intrinsic value of $Q_\mathrm{sg}$ in the dust and any assumed value when correcting for dust by starting from Equation \ref{Eqn: Eta Dust} and incorporating the definitions for $F_\mathrm{H\alpha, true}$ and $F_\mathrm{NUV,true}$, the ``true" flux emitted by the galaxy without any dust attenuation:

\begin{align}
\eta & =  \log_{10}\left( \frac{C_1 F_\mathrm{H\alpha,obs}}{C_2 F_\mathrm{NUV,obs}} \right) + \nonumber \\ & \hspace{0.8cm} 0.4\left[ \frac{k_\mathrm{H\alpha}}{Q_\mathrm{sg,out}} - k_\mathrm{NUV} \right]E(B-V)_\mathrm{stars}
\end{align}
\begin{align}
F_\mathrm{H\alpha,obs} & = F_\mathrm{H\alpha,true}10^{-0.4k_\mathrm{H\alpha}\,E(B-V)_\mathrm{stars}/Q_\mathrm{sg,in}}\\
F_\mathrm{NUV,obs} & = F_\mathrm{NUV,true}10^{-0.4k_\mathrm{NUV}E(B-V)_\mathrm{stars}}
\end{align}
\begin{align}
\eta & = \log_{10}\left( \frac{C_1 F_\mathrm{H\alpha,true}}{C_2 F_\mathrm{NUV,true}} \right) + \nonumber \\ & \hspace{0.8cm} 0.4 \left[ \frac{k_\mathrm{H\alpha}}{Q_\mathrm{sg,out}} - \frac{k_\mathrm{H\alpha}}{Q_\mathrm{sg,in}} \right] E(B-V)_\mathrm{stars}\\
\eta & \propto 0.4k_\mathrm{H\alpha}\left[ \frac{1}{Q_\mathrm{sg,out}} - \frac{1}{Q_\mathrm{sg,in}} \right] E(B-V)_\mathrm{stars} \label{Eqn: eta Rstar}
\end{align}
Here, $Q_\mathrm{sg,out}$ represents the value of $Q_\mathrm{sg}$ we assume when correcting for dust.  For the dust-corrected H$\alpha$ fluxes in the 3D-HST sample, $Q_\mathrm{sg,out} = 0.44$.  $Q_\mathrm{sg,in}$ represents the value of $Q_\mathrm{sg}$ assumed when adding dust effects to the ``true'' fluxes.  For $Q_\mathrm{sg,in} = 1.0$ and $Q_\mathrm{sg,out} = 0.44$, Equation \ref{Eqn: eta Rstar} gives $\eta \propto 1.69 E(B-V)_\mathrm{stars}$ (with the opposite slope flipping the input and output values).  These slopes agree well with the measured slopes in the top middle and lower left panels of Figure \ref{Fig: eta dust sigmaclip}.

The right two panels in Figure \ref{Fig: eta dust sigmaclip} show the 3D-HST sample for $Q_\mathrm{sg}=0.44$ and $1.0$.  The trend shown in the upper right panel is similar to the one in Figure \ref{Fig: Burstiness HST}, which is expected due to the positive correlation of $E(B-V)_\mathrm{stars}$ with $M_*$.  Naively, the slope shown in the top right panel of Figure \ref{Fig: eta dust sigmaclip} and in Figure \ref{Fig: Burstiness HST} seems to indicate that effects are present in the 3D-HST sample which we have not accounted for in the mock catalogs.  We have shown in Section \ref{Sec: Dust} that an incorrect measurement of $E(B-V)_\mathrm{stars}$ does not have a significant effect on $\eta$ and as a result, poor dust measurement \deleted{could}\added{should} not have produced such a strong relationship.  Instead, we can see from the top left and middle bottom panels in Figure \ref{Fig: eta dust sigmaclip} that when we implement dust attenuation in the mock catalog with a given value of $Q_\mathrm{sg}$, the distribution of $\eta$ is not greatly affected if we assume the same value of $Q_\mathrm{sg}$ when correcting for the dust later.  However, adding dust with a given value of $Q_\mathrm{sg}$ and correcting for it with the incorrect value leads to a strong relationship between $\eta$ and $E(B-V)_\mathrm{stars}$, as we see in the bottom left and top middle panels.  We can tune the value of $Q_\mathrm{sg}$ \added{used to correct the 3D-HST data }until we measure zero slope \deleted{to find the value implied by the 3D-HST data}\added{as predicted in the mocks}.  We find that $Q_\mathrm{sg} = 0.8^{+0.1}_{-0.2}$ produces $\eta = 0.00 E(B-V)_\mathrm{stars} - 0.51$ with the error range approximated by eye because the formal error bars obtained from Monte Carlo realizations appear to underestimate the true uncertainties.  For a galaxy population such as this 3D-HST sample ($0.65<z<1.5$), this could be an indicator of a transition from the $Q_\mathrm{sg}=0.44$ regime at low redshifts to the $Q_\mathrm{sg}=1.0$ regime closer to $z\sim2$.  

Applying $Q_\mathrm{sg} = 0.8^{+0.1}_{-0.2}$ to the relationship between $\eta$ and $M_*$ for the 3D-HST sample reveals that this value of $Q_\mathrm{sg}$ causes $\eta$ and $M_*$ to become\added{ nearly} independent as well.  Figure \ref{Fig: eta mstar} shows that the slope is reduced by a factor of $\sim5$.  Additionally, plotting the SFRF for the 3D-HST sample of galaxies using this measurement of $Q_\mathrm{sg}$ gives much better agreement both with \cite{Sobral2014} and the SAM and MUFASA simulations (Figure \ref{Fig: SFRF} right panel, orange dashed line).  \deleted{This lends credence to the possibility of improper assumption}\added{Hence, changing the assumed value} of $Q_\mathrm{sg}$ for the 3D-HST sample\deleted{, because this one change} can resolve multiple sources of tension between 3D-HST and the mocks.  We also note that when calculating the distribution of values after removing the best-fit trend shown in Figure \ref{Fig: eta mstar}, there are no significant changes to \added{the }$\sigma_\eta$ or $\gamma_\eta$ \deleted{as}\added{values} reported in Table \ref{Tbl: mockstats}.

However, it is important to note that $Q_\mathrm{sg}=0.8$ gives $\bar{\eta} = -0.31$.  This would indicate a strongly declining average SFH in the 3D-HST sample in the past $\sim100$\,Myr, which is unlikely for an emission-line-selected sample.  If we consider the lower bound of $Q_\mathrm{sg}=0.6$, we still find that $\bar{\eta} = -0.13$, implying a declining SFH.  As a result, it is not possible to find a value of $Q_\mathrm{sg}$ that resolves the dependence of $\eta$ on $E(B-V)_\mathrm{stars}$ and $M_*$ without also \deleted{measuring}\added{implying} a declining SFH in the 3D-HST sample.  This issue is not alleviated by adopting a different dust law because, since our analysis only utilizes two wavelengths, it is sensitive only to the ratio of dust attenuation between those two, which is equivalent to varying $Q_\mathrm{sg}$.  While varying $Q_\mathrm{sg}$ cannot make the 3D-HST sample fully consistent with the mocks, it does significantly reduce the tension between them, motivating further investigation into the dust properties of the 3D-HST galaxies.

\section{Conclusions}\label{Sec: Conclusions}

As of yet, there has been no \deleted{consistent} method of quantifying galaxy population burstiness consistently used in the growing number of papers on the topic.  Here, we attempted to give a full treatment to \deleted{galaxy star-formation} burstiness by starting from \added{individual galaxies' }intrinsic SFRs\deleted{,}\added{ and} forming a theoretical burst indicator ($\eta_\mathrm{th}$).  This theoretical burst indicator differentiates between periods of rising, falling, and mostly continuous SFRs.

We translated $\eta_\mathrm{th}$ into an observationally measurable value, $\eta$, by using \added{the time-dependent H$\alpha$ and NUV luminosities}\deleted{FSPS luminosity responses for the UV and H$\alpha$ star-formation rates}.  We calculated the detailed \added{time dependence of the H$\alpha$ and NUV luminosities}\deleted{temporal luminosity responses for showed that $\mathrm{SFR_{H\alpha}}$ and SFR$_\mathrm{NUV}$}, and found that the time for each to reach [10\%, 50\%, 90\%] of the maximum luminosity when undergoing continuous star formation is [1.32, 2.58, 5.03] Myr for H$\alpha$ and [2.87, 24.56, 520.64] Myr for the NUV respectively.  We use the average, standard deviation, and skewness of the distribution of $\eta$ to \deleted{determine}\added{describe} the properties of the population as a whole.  The average and skewness can indicate trends in the recent SFR of the population, and the standard deviation is an indicator of the total variability (burstiness) in SFR.

We analyzed the effects of changing the high-mass IMF slope and metallicity of the galaxy sample on the distribution of $\eta$.  We found that extreme changes to the high-mass IMF slope can moderately enhance or suppress burstiness measurements, while extreme changes to metallicity did not induce significant changes in the width of the distribution of $\eta$.  Neither IMF slope nor metallicity has a strong effect on $\bar{\eta}$.  Additionally, when examining the effects of bulk trends in SFR, we found that galaxies \deleted{which}\added{that} evolve along the SFR-M$_*$ relation will cause a small ($\lesssim0.1$) increase in $\bar{\eta}$ and that only galaxies with an average SFH that is rising or falling rapidly affect $\bar{\eta}$.  Our investigations imply that $\bar{\eta}$ (or the average of similar indicators), although robust to some systematics, is a better probe of average SFH or dust law errors than of burstiness.

We then applied realistic noise and selection effects on NUV and H$\alpha$ measurements, generating mock catalogs of H$\alpha$ and NUV fluxes from the SAM and {\sc Mufasa} simulated star formation rates at \added{3D-}HST and JWST depths.  We discovered that there is potentially a systematic bias in the 3D-HST dust corrections \deleted{which}\added{that} leads to a disagreement in SFRF \deleted{both }with\added{ both} the SAM and {\sc Mufasa} simulated galaxies and with the results of \cite{Sobral2014}.  However, under the assumption that $Q_\mathrm{sg} = E(B-V)_\mathrm{stars}/E(B-V)_\mathrm{gas} = 0.44$ using a \cite{Calzetti2000} dust law, $\eta$ is robust to incorrect measurements of dust, so we \deleted{were able to }continue\added{d} our analysis. 

The width of the $\eta$ distribution is wider for the 3D-HST sample than that of the mock catalog, which could be attributed either to genuinely greater burstiness in the 3D-HST sample or to an unknown source of scatter \deleted{which}\added{that} was not present in the mocks.  We also found a trend between $\eta$ and $M_*$ and between $\eta$ and $E(B-V)_\mathrm{stars}$\deleted{ that}\added{, which} is not present in the mocks.  Because these trends could not be caused by improper dust measurement, we investigated the effects of changing $Q_\mathrm{sg}$, \added{which }modulat\deleted{ing}\added{es} the relative attenuation effectiveness of dust between nebular and stellar emission.  Because our analysis incorporates only two measurements of SFR, varying $Q_\mathrm{sg}$ effectively allows us to probe the full range of effect\deleted{ on our analysis inherent in allowing for }\added{ of }different dust prescriptions.  We found that we could remove the dependence of $\eta$ on $E(B-V)_\mathrm{stars}$ and $M_*$ while also resolving the SFRF discrepancy in the 3D-HST sample by using $Q_\mathrm{sg} = 0.8^{+0.1}_{-0.2}$, which is between the values assumed at low redshift ($Q_\mathrm{sg} =0.44$) and some findings at high redshift ($Q_\mathrm{sg} =1.0$).  The drawback to this approach was that we were left with $\bar{\eta}<0$ for the full range of possible values of $Q_\mathrm{sg}$ which resolve the discrepancy, implying a declining SFH that is in disagreement with the mocks\deleted{,} and \deleted{that is }unlikely for a sample selected as strong H$\alpha$ emitters.

Because of this enduring discrepancy and because the 3D-HST mocks indicate that measurements of $\eta$ are\deleted{ greatly} affected by noise, we look forward to the improvements in accuracy JWST will bring to H$\alpha$ measurements.  Our mocks that incorporate estimated JWST selection effects find that JWST will greatly improve the measurement of \deleted{intrinsic }burstiness for galaxies at $z\sim1$.  Another approach for improving burstiness measurements of the 3D-HST sample would be to follow up with infrared observations of these galaxies.  Measurements of SFR$_\mathrm{FIR}$ would help to resolve the degeneracy between dust prescription assumptions and increased burstiness in 3D-HST while also potentially alleviating tension when comparing the 3D-HST SFRF to that of other data sets.

\section*{Acknowledgements}
The authors would like to thank Christopher Hayward for his insightful comments and suggestions, and they gratefully  acknowledge  support
from Rutgers University.  The  Flatiron Institute is supported by the Simons Foundation.  Support for Program number HST-AR-14564.001-A was provided by NASA through a grant from the Space Telescope Science Institute, which is operated by the Association of Universities for Research in Astronomy, Incorporated, under NASA contract NAS5-26555.

\bibliographystyle{apj_url}
\bibliography{bibliography}

\end{document}